\documentclass[fleqn,usenatbib,useAMS]{mnras}

\usepackage{graphicx} % Including figure files
\usepackage[fleqn]{amsmath} % Advanced maths commands
\usepackage{amssymb} % Extra maths symbols
\usepackage{multicol}% Multi-column entries in tables
\usepackage{multirow}
\usepackage{array}
\usepackage{bm} % Bold maths symbols, including uprightGreek
\usepackage{pdflscape,float} % Landscape pages
\usepackage{natbib}

\hypersetup{final}
\hypersetup{colorlinks,citecolor=blue}
\begin{document}
%==============================================================================================

%==============================================================================================
\title[Are superthins edge-on LSBs?]
{Are superthin galaxies low surface brightness galaxies seen edge-on? The star formation probe}
\author[Ganesh Narayanan \& Arunima Banerjee]
       {Ganesh Narayanan$^{1}$\thanks{E-mail: ganeshn@students.iisertirupati.ac.in} and 
        Arunima Banerjee$^{2}$\thanks{E-mail :  arunima@iisertirupati.ac.in} \\
\{$1$, $2$\}  Indian Institute of Science Education and Research, Tirupati 517507, India \\ 
} 
\maketitle

\begin{abstract}
Superthin galaxies (STs) are edge-on disc galaxies with strikingly high planar-to-vertical axes ratios of $\sim 10 - 20$ with no bulge component, and central surface brightness in $B$-band $>$ 23 mag arcsec$^{-2}$ comparable to low surface brightness galaxies (LSBs).  Although STs and LSBs have similar dynamical, stellar and atomic hydrogen (HI) masses on an average, it is tricky to conclude if they constitute the same galaxy population, given the edge-on and face-on orientations of the STs and the LSBs respectively. We systematically study star formation rate (SFR) in a sample of STs and LSBs using SED fitting of photometric data in ten bands including GALEX: FUV, NUV, SDSS: u,g,r,i,z \& 2MASS: J, H, Ks using stellar population synthesis models employing the publicly-available software MAGPHYS (Multi-Wavelength Analysis of Galaxy Physical Properties). The estimated median SFRs for LSBs and STs are 
$0.4^{+2.2}_{-0.3} $ $M_{\odot}yr^{-1}$ \& $0.2^{+0.9}_{-0.2}$ $M_{\odot}yr^{-1}$ respectively. Our calculations indicate that this deficit in the SFR of an ST can be attributed to inclination and opacity effects. Therefore, we conclude that STs and LSBs have equal intrinsic SFR over and above other physical properties, which possibly implies that STs are just LSBs seen in edge-on.
\end{abstract}

\begin{keywords}
galaxies: star formation, galaxy:  evolution, ultraviolet: galaxies,  galaxies: spiral,  
galaxies: disc, ISM: dust, extinction
\end{keywords}

\section{INTRODUCTION}
%======================================================================================================================
\noindent Superthin galaxies (STs) are bulgeless edge-on disc galaxies with strikingly high planar-to-vertical axes ratio ($a/b \sim 10 - 20$) \citep{1999astro.ph.11022M}.  The origin of these superthin stellar discs continues to be a puzzle, but is crucial to understand the characteristic structure and kinematics of these galaxies. Besides, galaxy interactions are not rare and may lead to heating up and thickening of the stellar discs \citep{Walker96,Velazquez99,QuY11,Mihos1996}. Independent of the drivers of thick disc formation, most of the large galaxies are now known to host a thick disc \citep{GilmoreR1983,Dalcanton2002}. Therefore, the existence of low surface brightness with a superthin stellar vertical structure with no discernible bulge component is a mystery.  A superthin stellar disc indicates the presence of an ultracold stellar disc i.e., a stellar disc with strikingly low values of stellar vertical velocity dispersion \citep{Aditya2021}. The existence of ultra-cold stellar discs is an enigma, given the understanding that galaxies form via mergers in the hierarchical structure formation scenario, However, galactic discs are liable to be heated by internal agents like non-axisymmetric instabilities (bars and spiral arms) and other routes of secular evolution \citep{barbanis1967, Grand2016}. This possibly indicates that STs are lacking in strong bars and spiral arms, as is expected in very late-type, under-evolved systems However, the absence or near-absence of bars and spiral arms in highly edge-on systems like the STs is not amenable to observational confirmation of (But see, \cite{Bureau2005}).

These apart, STs are fascinating for an additional reason; all such galaxies studied till date are low-surface-brightness (LSB) galaxies \citep{Goad1981,Mathews2000} although the reverse is not true (\cite{Bizyaev_2004}, table 3). Low Surface Brightness galaxies (LSBs), in turn, are face-on galaxies with central $B$-band surface brightness $\mu_{B} >$ 23 mag arcsec$^{-2}$ \citep{1997PASP..109..745B,2001AJ....121.2420S}, or alternatively, with central $R$-band surface brightness $\mu_{B} >$  24 mag arcsec$^{-2}$ \citep{1996ApJS..103..363C, Adami_2006}. They are gas-rich and dark matter dominated with low SFRs, and therefore constitute proxies for a primeval galaxy population in the local universe \citep{1993egte.conf...92V,1997PASP..109..745B,1994Natur.367..538M,10.1093/mnras/274.1.235,1993egte.conf...92V,2000A&A...357..397V,Bell2000}. The LSBs are common, are highly gas rich yet do not appear to have had a significant star formation, and their origin and dynamics are not understood well \citep{de_Blok_2005}. Interestingly, LSBs are found to be located at the outskirts of voids, which, therefore, which may explain the quiescent nature of their stellar discs \citep{Rosenbaum2009}. Thus, understanding why some LSBs are superthin may help shed light on the overall evolution of disc LSB galaxies. Although STs and LSBs have almost comparable physical properties including dynamical mass, stellar mass and atomic hydrogen mass, it is tricky to conclude if they constitute one and the same galaxy population, given the ambiguity in modelling the surface densities of STs and vertical scale heights of LSBs due to their edge-on and face-on orientations respectively. 

Earlier studies indicate that the average SFRs of LSBs  is lower by at least an order of magnitude than HSBs \citep{1997PASP..109..745B}. Further these galaxies are bluer in color ($U - B < 0$) when compared to high surface brightness galaxies (HSBs) \citep{10.1093/mnras/274.1.235,1994AJ....107..530M}, possibly indicating a predominantly young stellar population \citep{10.1046/j.1365-8711.1999.02282.x,10.1093/mnras/274.1.235}. 
This is also in compliance with the fact that these galaxies are low mass galaxies ($M_* > 10^8 M_{\odot}$), constituting 47 \% of local population when compared to typical HSBs \citep{Martin2019}. However, the age-metallicity degeneracy poses a challenge in determining age from galaxy colors \citep{Worthey1994}. In any case, from bluer optical colors of LSBs, \cite{10.1093/mnras/274.1.235} ruled out various scenarios such as a disc-fading, an initial starburst, an exponentially declining or a constant SFR in the star formation history (SFH) of LSBs. \cite{2000A&A...357..397V} in turn showed that LSBs have an exponentially decreasing SFRs, and follow same evolutionary history as HSBs, but at a slower rate with only a few star formation bursts is required to explain blue color of LSBs. However, there exists a large fraction of population of red LSBs which do not fit into the proposed model for the formation of blue LSBs \citep{O_Neil_2008}. Besides, current SFRs of LSBs are generally higher than their past SFRs, which is again indicative of a young stellar population, and which may possibly be attributed to a late epoch of formation or a  slow paced evolution. A recent study with $H_{\alpha}$ luminosity for 357 LSBs showed that the sample has a lower global SFR ($  \rm{log}(SFR)(M_{\odot} yr^{-1}) = -1.5$) and lower star formation density ($ \rm{log}(\Sigma_{SFR})(M_{\odot} yr^{-1} kpc^{-2}) = -3.3 $) relative to HSBs \citep{Lei_2019}; $H_{\alpha}$ traces the SFR in the last 3-10 Myrs. 

\cite{Melnyk_2017} compared and contrasted the star-formation properties of superthins (Ultra Flat Galaxies -UFGs ) with major-to-minor axes ratio $(a/b)_B>10$ and non-superthins from the RFGC, and found that the specific star formation rate $sSFR$ as traced by GALEX FUV  increases steadily from the early-type to late-type discs for both the samples, without any striking difference between the different morphological types. Further, the population of superthin (UFG) discs has the average HI mass-to-stellar-mass ratio by (0.25 ± 0.03) dex higher than that of RFGC galaxies in general, with a specific star formation rate (sSFR) $\sim$ -10.56 (in $\rm{log}10{M_{\odot}{\rm{yr}}^{-1}}$ units, the average star formation rate being approximately three times the current SFR. They finally concluded that the UFG galaxies have also sufficient amount of gas to support their observed SFR over the following nearly 9 Gyrs. Further \cite{Karachentseva_2020} used GALEX FUV luminosity to determine the SFR for 181 nearly face-on spiral galaxies ($\rm{log}(a/b) < 0.05 $) with median asymptotic rotational velocity around $60 \rm{km s}^{-1}$, assumed to be thin galaxies, from HyperLeda database\citep{Markrov2014}. They found that the average value of the sSFR for these galaxies falls off smoothly from low-mass to giant discs, and that the SFR in the past must be two to three times higher than the current SFR to match the present stellar mass. 

In this paper, we compare and contrast the SFRs of STs and LSBs in a systematic manner. We consider the STs from the RFGC \cite{karachentsev2003revised} and LSBs from \cite{refId0}. The median asymptotic velocities of our sample STs and LSBs are 131 $\rm{kms}^{-1}$ and 114 $\rm{kms}^{-1}$, and hence they constitute intermediate mass galaxies unlike the sample studied by \cite{Melnyk_2017} with a mean asymptotic velocity of $\sim$ 60 $\rm{kms}^{-1}$.
We estimate the SFR using FUV luminosity from GALEX \citep{Bianchi_2017} \& NIR from WISE W3 \citep{2012yCat.2311....0C} for a sample of STs and LSBs. FUV traces the emission from young stellar populations (100 Myrs) and estimates the recent SFR in the galaxy which however is highly sensitive to dust extinction. WISE NIR similarly gives SFR by tracing a young stellar population (100 Myr) but is relatively unaffected by dust extinction effects. 
Besides, we carry out SED fitting of photometric data in ten bands including GALEX: FUV, NUV, SDSS: u,g,r,i,z \& 2MASS: J, H, Ks using stellar population synthesis models with MAGPHYS. Our estimated SFRs for STs and LSBs in the NIR band are 0.5 and 0.2 M${\odot}$yr$^{-1}$ respectively. Finally, we also estimate the apparent SFR of an ST when viewed face-on, taking into account the effect of inclination and opacity corrections for a few STs from our sample. 
The rest of the paper is organized as follows. In \S 2, we present the theory, in \S 3, we describe our sample, in \S 4, we present our results followed by discussion and conclusions in \S 5 and 6 respectively.

\section{Theory}
\subsection{Observational Tracers of Star Formation}
\subsubsection{ GALEX (FUV) and WISE W3 (NIR)  }

Due to the deep sensitivity and wide field-of-view, GALEX observations are suitable for studying star formation properties of outer galactic discs, tidal tails, LSBs and dwarf Irregular
galaxies \citep{2011Ap&SS.335...51B}. Young massive stars are tracers for recent star formation activities in the galaxy occurred for the last $\sim$ 100 Myrs \citep{2011ApJ...737...67M,2011ApJ...741..124H}. 
$SFR / L_{u}$ varies over a magnitude for range of galaxy colors, $L_u$ being the luminosity in the $u$-band.  In UV, SFR scales linearly with luminosity dominated by spectrum of young stars as given by

\begin{equation}
SFR(M_{\odot}yr^-1)=1.4* 10^{-26} L_{\nu} (erg s^ {-1} Hz^{-1})
\end{equation}

\noindent \citep{Kennicutt1998}. In deriving the above relation, it was assumed that the SFR remained constant for time scale larger compared to age of dominant UV emitting stars in continuous star formation approximation. The IMF(Initial mass function) chosen was the Salpeter IMF. This relation can only be used when the SFR is constant over timescales $> 10^8 yr$ and is also sensitive to the choice of Initial Mass Function (IMF) and extinction. \\

 However, UV is highly sensitive to dust extinction and any corrections for extinction in UV gives a large scatter in the SFR relation \citep{Calzetti_2007}. Besides, \cite{deJong1996A&A} showed that such degeneracy could be partially broken by adding NIR photometry to optical colors, which has also been stated by \cite{Bell2000} and \cite{Wu_2005}. \cite{2005A&A...432..423T} showed  that  the  $12\mu m $ luminosity  could  be  used  as  a reliable measure of total infrared luminosity and can be used for calibrating SFR. \cite{Cluver_2017} gives the SFR correlation with WISE luminosity. This equation was a linear fit to SFR and total infrared luminosity for galaxies from SINGS and KINGFISH using Kroupa's IMF for calculations. The infrared luminosity traces a SFR over last 100 Myr which sensitive to star formation history \citep{2011ApJ...737...67M,2011ApJ...741..124H}.

%\begin{multline*}
\begin{equation}
\rm{log} SFR (M_{\odot}yr^{-1}) = (0.889 \pm 0.018)log  L_{12 \mu m}(L_{\odot})\\
-(7.76 \pm 0.15)
\end{equation}
%\end{multline*}
%$$
%\eqno(2)
%$$

The SFRs based on the Salpeter IMF are known to be larger than those based on other IMFs such as Kroupa and Chabrier by a factor of 1.4 - 1.6 e.g., \citep{Calzetti_2007,Kennicutt_2009}.  Therefore, it is necessary to take into account these offsets when one compares the calibration results with different IMFs. WISE (W4)  band  is not  contaminated by emission lines  (z=0) and so in the absence of active galactic nucleus (AGN) activity, gives us a reliable measure of star formation comparable  to $H\alpha$ measures \citep{Brown_2017}. However,  the  widespread  use  of this band is severely hampered due to sensitivity issues \citep{2013AJ....145....6J}. 
SFR luminosity correlation in W4 band is given as below. The SFR is correlated to SFR estimated in W3 band.
\begin{equation}
\rm{log} SFR (M_{\odot}yr^{-1}) = (0.915 \pm 0.023)log  L_{23 \mu m}(L_{\odot})\\
-(8.01 \pm 0.20)
\end{equation}

\subsubsection{SED fitting using MAGPHYS} 

Theoretical modelling of SED is done by solving radiative transfer equations for a known stellar population and other components. Stellar population synthesis models returns the temperature and bolometric luminosity for various stellar masses as a function of time for simple stellar populations and the integrated flux from the grid is obtained from stellar spectral libraries. Stellar templates are added with IMF to obtain stellar spectra for same age population of stars. These isochrones are taken linear combination to obtain the final spectra with an exponentially parameterized star formation history. Thus this population synthesis modelling contains minimum of four parameters such as IMF, star formation history, age of the stellar population and metallicity.

In this study we use publicly available software MAGPHYS (Multi-wavelength Analysis of Galaxy Physical Properties) for SED fitting \citep{2008MNRAS.388.1595D}.
The code uses a stellar population synthesis code by \cite{2003MNRAS.344.1000B} to predict the spectral evolution of stellar population. The dust attenuation of stellar spectrum is modelled using the two component model of \citet{2000ApJ...539..718C}.
The specific intensity of the stellar population at time t with a star formation rate $\psi(t)$ is given as

\begin{equation}
L_{\lambda}=\int \psi(t') S_{\lambda}(t',Z) e^{-\tau_{\lambda}}dt'
\end{equation}

Here Z is the metallicity of the population and $\tau_{\lambda}$ is the opacity of ISM, $S_{\lambda}$ is the specific intensity of the simple stellar population (SSP) of age t'. SFR is again modelled as continuous function parameterized by 
$t_{form}$ and time scale parameter $\gamma$. Thus $\psi(t)\propto e^{- \gamma t}$ and random bursts are superimposed to this continuous model.

The optical depth is modelled according \cite{2000ApJ...539..718C}. The attenuation for the young stars and old stars is different so that 
\\
\begin{equation}
\tau_{\lambda} = \left\{
                \begin{array}{ll} \tau^{BC}+\tau^{ISM} \hspace{5mm} t' \leq 10^7 yr\\
                \tau^{ISM} \hspace{15mm} {t' > 10^7 yr}\\
         \end{array}
              \right.
\end{equation}
The spectral energy distribution of the power reradiated by dust in the stellar birth clouds is computed as the sum of three components: polycylic aromatic hydrocarbons (PAHs), mid-infrared continuum characterizing the emission from hot grains, and grains in thermal equilibrium with adjustable temperature \citep{2008MNRAS.388.1595D}. 

The optical broadband colors have the intrinsic problem of age-metallicity degeneracy \citep{Worthey1994,MacArthur_2004}. FUV and NUV photometry with optical observations in the standard broad bands help to break the age-metallicity degeneracy. Therefore, we emphasise that the star formation rate estimated by SED fitting using MAGPHYS in ten photometric bands (See \S 3) is the most appropriate measurement of our analysis in this paper. However, in addition, we have estimated the SFR using the GALEX FUV and the WISE W3 NIR tracer. While both trace the same stellar population giving the average SFR in the last 100 Myrs, the former is prone to dust extinction unlike the latter (\S 2.1.1). Therefore, comparison of SFR estimation from GALEX and  WISE may indirectly help assess the dust extinction effect in the galaxies, as is seen in our study (\S 4.1).
\subsection{Error Analysis} For the GALEX FUV (WISE W3 NIR) study, for each galaxy in the ST and the LSB sub-samples, we first determine the SFR using Equation 1(2) as detailed in \S 2.1.1. The error in the estimated SFR is then determined by propagating the errors in a) the magnitude and b) the distance to the galaxy. Next, for each such galaxy, we approximated the SFR distribution as a Gaussian with the mean equal to the SFR, and the standard deviation equal to the error in the SFR as estimated above. We repeated the above process for all the galaxies in the LSB and ST sub-samples separately for the GALEX FUV (WISE W3 NIR) study. Thus we get the SFR distribution of each galaxy in the LSB and ST sub-samples in the GALEX FUV (WISE W3 NIR) study. For the SED fitting study, MAGPHYS directly outputs the likelihood distribution of the SFR values for each galaxy. Finally, we do a random (Monte-Carlo) sampling of about $\sim$ 100 data points from the SFR distribution of each galaxy in the ST (LSB) sample for each of the above studies (GALEX, WISE, MAGPHYS). For each study, the data points from all the STs are then combined to get the SFR distribution of the \emph{entire} ST (LSB) sub-sample. The median of this distribution is now quoted as the representative SFR of ST (LSB) from the study; the difference of the median from the first and the third quantiles represent the error bars. The same technique was adopted for obtaining the representative SFR and the error bars for the LSB sub-sample for a given study.
    
\begin{figure*}
\begin{tabular}{ccccc}
\resizebox{65mm}{!}{\includegraphics{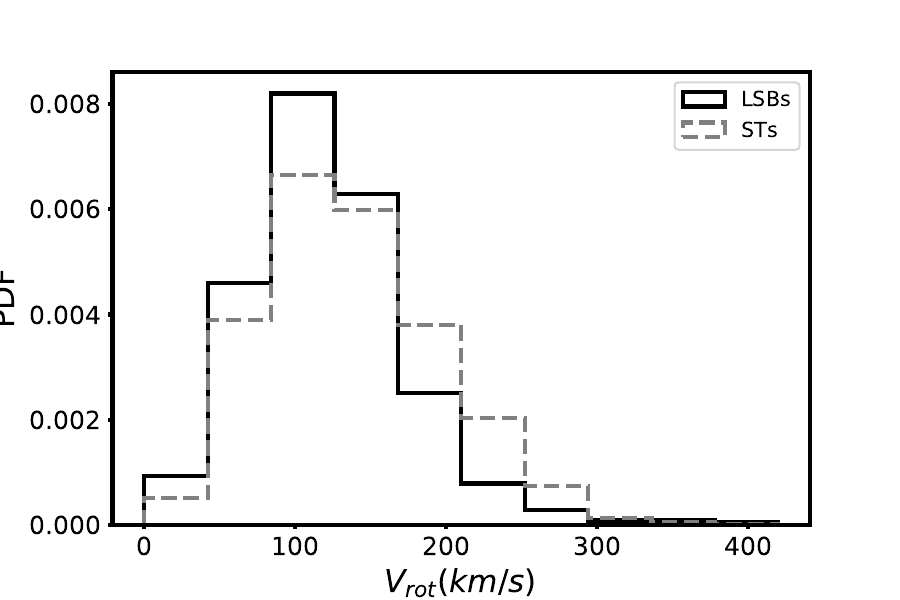}}
\resizebox{65mm}{!}{\includegraphics{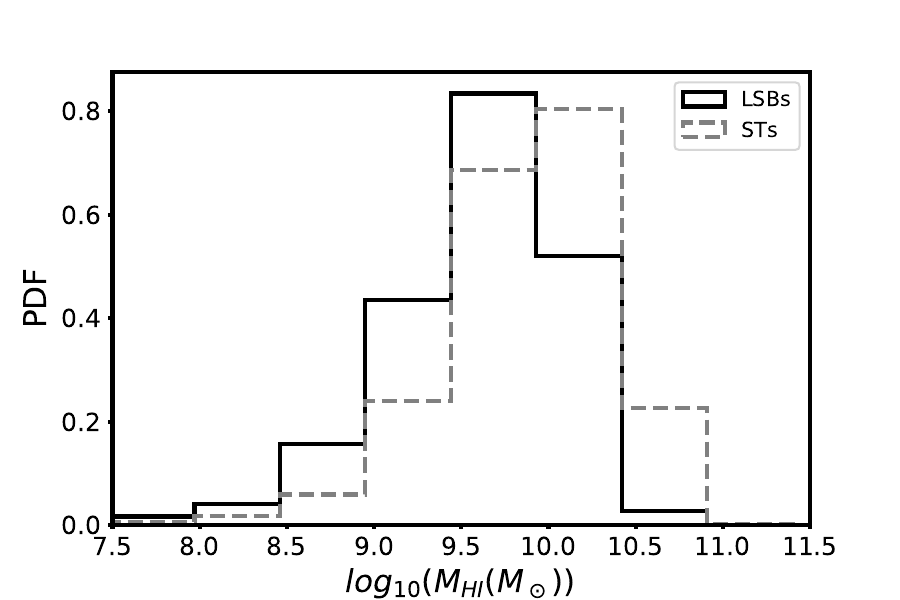}} \\
\resizebox{65mm}{!}{\includegraphics{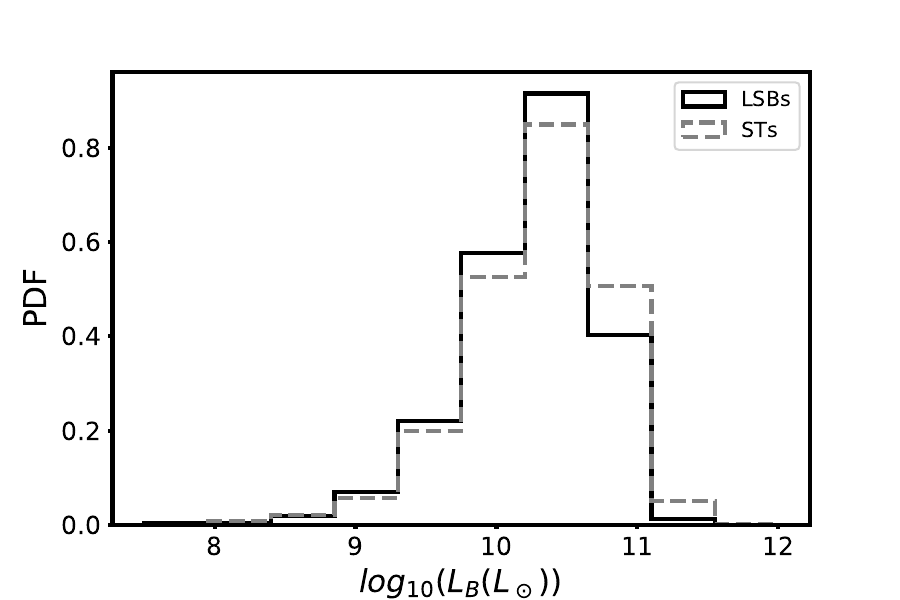}}
\resizebox{65mm}{!}{\includegraphics{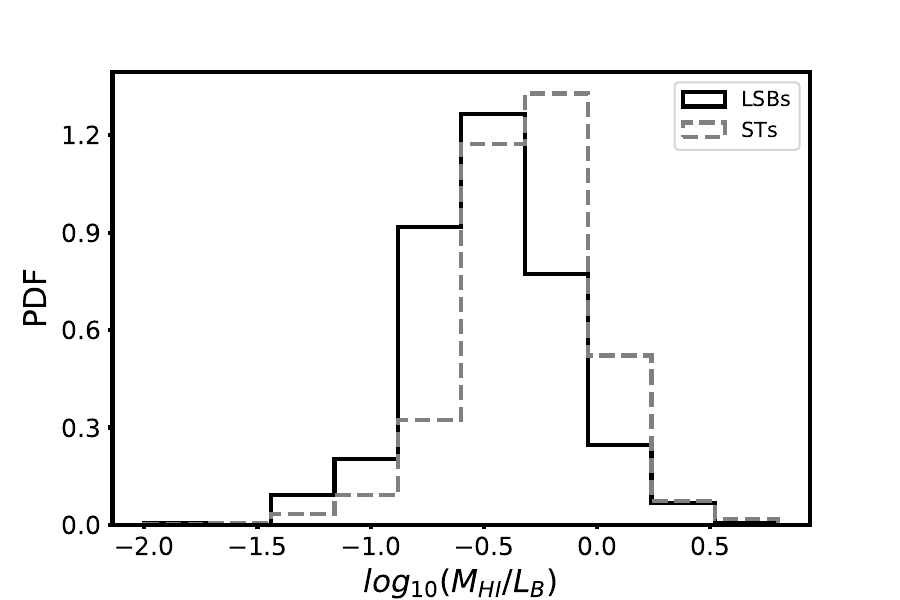}} \\
\resizebox{65mm}{!}{\includegraphics{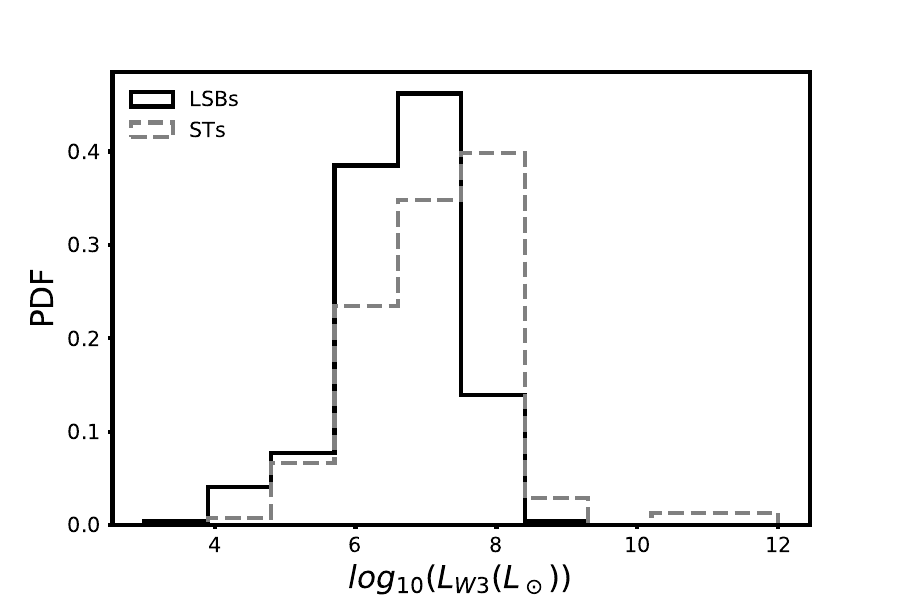}} 
\resizebox{65mm}{!}{\includegraphics{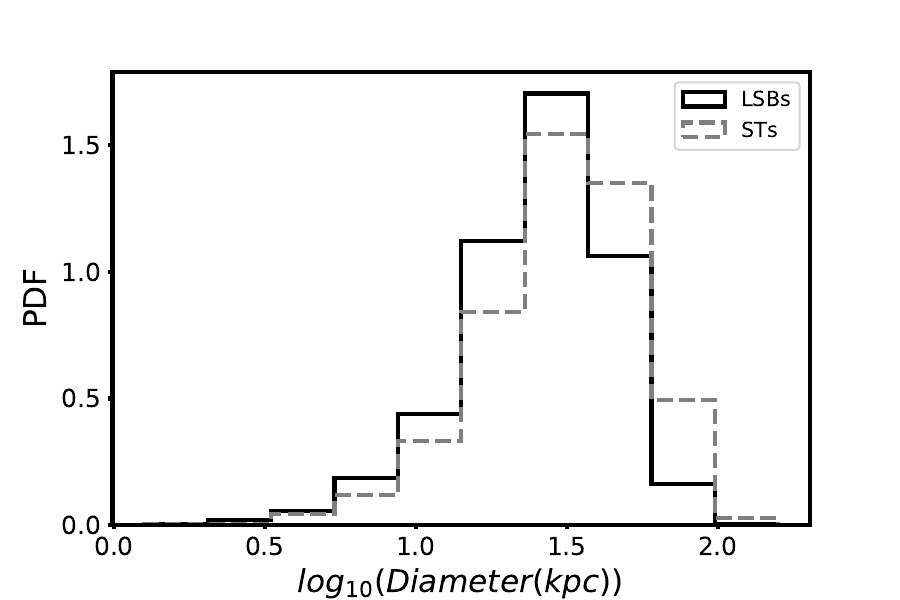}}\\ 
\end{tabular}
\caption{Top Panel [Left] Distribution of asymptotic rotational velocity $V_{\rm{rot}}$ ($\rm{kms}^{-1}$) [Right] Distribution of atomic hydrogen mass $\rm{log(M_{HI}/M_{\odot}})$. Middle Panel [Left] Distribution of B-Band luminosity $(\rm{L_B/L_{\odot}})$ [Right] Distribution of $\rm{log(M_{HI}/L_B})$ . [Bottom Panel] Left: Distribution of $log(L_{W3}/L_{\odot})$ in WISE(W3) band [Right] Distribution of diameter in B band. }
\label{A1.Physical properties}
\end{figure*}

\section{SAMPLE}
We have chosen the LSBs sample from the NIR catalogue of LSBs of \cite{refId0}. The STs sample was chosen from \cite{karachentsev2003revised}. These were cross matched with the catalogues of 
1) \cite{Bianchi_2017} to obtain the GALEX FUV and NUV magnitudes of our sample LSBs and STs.  
2) \cite{2012yCat.2311....0C} to obtain the WISE W3 $12 \mu$ data. Only those galaxies were chosen for which distances were available in the NED database as was required for the conversion of magnitude to luminosity. For SED fitting using MAGPHYS, we combine  GALEX (NUV, FUV), SDSS (u,g,r,i,z)\citep{AhnSDSS2012} and NIR (J,H,Ks) photometric data. The photometric redshift was obtained from NED database. A cross match between \cite{refId0} and \cite{karachentsev2003revised} gives zero overlap within 2 $\arcmin$ radius, indicating that our sample does not have LSBs which are also STs. Then those galaxies which gave $\chi^2 >5$ in SED fitting were rejected. The sample sizes are provided in Table 1. \\

 Star formation in galaxies is regulated by several physical parameters like asymptotic rotational velocity $(V_{\rm{rot}})$, the total stellar mass as traced by, for example, the WISE W3 band luminosity $(L_{W3})$, and the total atomic hydrogen mass $(M_{HI})$. In this study, we have checked if the distribution of physical properties associated with  star formation are well-matched for our ST and the LSB sub-samples. For each physical parameter, we first obtain the mean values for the ST and the LSB sub-sample respectively, and then the grand-mean of the combined ST + LSB sub-samples. Finally, we determine the standard deviations from the grand-mean for the ST and the LSB sub-sample respectively, and normalize them by the grand mean. We present our calculations for the GALEX FUV, WISE W3 NIR and MAGPHYS samples in Table 3 respectively. We note that for each study, and for each parameter, the mean values of the ST and the LSB sub-sample mostly match with each other and also with the grand-mean of the combined ST + LSB sub-samples. Further, the normalized standard deviation about the grand-mean for the ST and the LSB sub-samples match well. Therefore, we conclude, that our ST and LSB sub-samples are well-matched in the parameter space as far as their mean-values as well as the scatter about the mean values are concerned.\\

\noindent \textbf{Sample Properties:} 
Figure 1  shows the distributions of asymptotic rotational velocity $V_{\rm{rot}}$, total $HI$ mass $M_{HI}$, $B$-band luminosity $L_B$, $HI$ mass to blue luminosity ratio $M_{HI}/L_B$, total NIR luminosity in the WISE (W3) band $L_{\rm{W3}}$, and finally the $B$-band diameter for STs and LSBs from HyperLeda. STs  and LSBs are represented by grey and black colors respectively and the same colour convention is followed for rest of the plots. The median values of the above physical quantities for the STs and LSBs have been presented in Table 2.  Tables A1 and A2 list the $B$-band magnitude $M_B$, distance, asymptotic rotational velocity $V_{rot}$, axes ratios $a/b$, B-band central surface brightness ${\mu}_B$ and $B$-band diameter of the STs and LSBs samples. For the sake of comparison, we also provide the above properties of a subset of the sample of bulgeless disc galaxies from \cite{Karachentseva_2020} in Table A3. \\

\begin{table}
	\centering
	\caption{Star formation of LSBs and STs from different observational tracers: Sample sizes}
	\label{table:1}
	\begin{tabular}{lcc} % four columns, alignment for each
		\hline
		 Tracer & LSBs & STs \\ 
		
		\hline
		GALEX (FUV) & 212 & 158   \\
		  WISE W3 (NIR)  & 345 &  549 \\
		SED \ fitting (MAGPHYS) & 103 &  65\\
%		HYBRID (UV+IR) &201&136\\
		\hline
	\end{tabular}
\end{table}

\begin{table}
	\centering
	\caption{Physical properties: LSBs versus STs}
	\label{table:2}
	\begin{tabular}{lcc} % four columns, alignment for each
		\hline
		 Physical Quantity & LSBs & STs \\
		\hline
		Asymptotic rotational velocity $V_{\rm{rot}}$ (km/s)& 114&131  \\
		B-band Luminosity ($L_{\odot}$)&1.8E+10  & 1.7E+10 \\
		W3-band Luminosity $(\rm{log}_{10}(L_{W3}/L_{\odot}))$ & 6.7&7.3 \\
		Optical Diameter (kpc) &26.5 & 32.2 \\
		HI mass-to-B-luminosity $(\rm{log}_{10}(M_{HI}/L_B))$ &0.3 & 0.3 \\
	
		\hline
	\end{tabular}
\end{table}

\begin{table*}
	\centering
	\caption{ST and LSB sub-sample properties}
	\label{table:3}
	\begin{tabular}{lcccccc} % four columns, alignment for each
		\hline
		 Physical Quantity & LSBs & STs& Grand Mean &$\sigma / mean$ LSBs&$\sigma / mean$ STs\\
		\hline
		GALEX FUV: ST and LSB sub-sample properties\\
		\hline
		Asymptotic rotational velocity(km/s)& 106.9&106.5 &106.5&0.3 &0.2\\
		$log_{10}(L_{W3}(L_{\odot}))$ & 6.7&7.5&7.1 &0.04&0.05\\
		$log_{10}(M_{HI}(M_{\odot}))$  & 9.4&9.9&9.7&0.1&0.1\\
		\hline
	WISE NIR: ST and LSB sub-sample properties\\
		\hline
		
		Asymptotic rotational velocity(km/s)& 106.4 & 120.6 &113.6&0.3 &0.2\\
		$log_{10}(L_{W3}(L_{\odot}))$ & 6.5&7.5&7.0&0.04&0.03\\
		$log_{10}(M_{HI}(M_{\odot}))$  & 9.3&10.0&9.7&0.04&0.03\\
		\hline
	MAGPHYS: ST and LSB sub-sample properties\\
		\hline
		Asymptotic rotational velocity(km/s)& 115 & 115 &115.0 & 0.2 & 0.2\\
		$log_{10}(L_{W3}(L_{\odot}))$ & 7.8 & 7.39 & 7.6 & 0.05 & 0.05\\
		$log_{10}(M_{HI}(M_{\odot}))$  & 9.9 & 9.9 & 9.9& 0.1 & 0.1\\
		\hline
	\end{tabular}
\end{table*}

\section{Results}

\subsection{Star Formation from Observational Tracers: General Properties} 

\subsubsection{GALEX FUV \& WISE W3 NIR:} In Figure 2 (Top Left), we study the distribution of SFR as estimated from FUV magnitudes from GALEX for both STs and LSBs. We find the SFR is $0.2^{+1.9}_{-0.1}$ $M_{\odot} yr^{-1}$ for LSBs \& $0.1^{+1.8}_{-0.0}$ $M_{\odot} yr^{-1}$ for STs. These are the median SFR values with the "-" and the "+" signs indicating its difference from the first and the third quantiles respectively. We note that our estimated median SFR for STs  is lower than that of the LSBs. In contrast, \cite{Melnyk_2017} found the average SFR for 259 UFGs to be $\sim$ 0.4 $M_\odot / yr$ (corrected for dust extinction). The difference between our estimate and that of \cite{Melnyk_2017} may possibly be attributed to dust extinction due to edge-on geometry of the STs; the latter being prone to dust extinction and the GALEX FUV magnitudes for our sample galaxies were not corrected for internal dust extinction. 
However, our estimated SFR for LSBs is in compliance with the study of \cite{Wyder_2009}. They analysed star formation in a sample of 19 LSBs from \cite{10.1093/mnras/274.1.235}, with asymptotic rotational velocity $V_{rot}$  ranging between 29 - 214 $km/s$ and $r$ band magnitude $M_r$ between $-13> M_r >-23$, thus spanning galaxy dynamical masses from dwarfs to giants. Their results indicated SFRs in the range of 0.001 to 3.002 $M\odot yr^{-1}$ with 0.27 $M\odot yr^{-1}$ as median where the lowest and highest SFR corresponds to a dwarf Irregular (dI) and a giant LSB (GLSB) respectively. 
We note the SFR estimate from FUV is more reliable than that from NUV as the calibration between SFR and UV luminosity was not done considering the entire range of NUV ($1750-2800 \mathrm{\AA}$) \citep{Roychowdhury_2009}. Figure 2 (Top Right) shows the distribution of SFR estimated from WISE W3 luminosity for sample of LSBs and STs. It is $0.3^{+2.1}_{-0.1} $ $M\odot yr^{-1}$ for LSBs \& $0.3^{+2.2}_{-0.1}$ $M\odot yr^{-1}$ for STs. We note, in the WISE W3 band, which is relatively free from dust-extinction effects, the median SFR values of STs and LSBs are equal.

\subsubsection{SED Fitting using MAGPHYS:} 

\begin{table}
\begin{minipage}{10cm}
{
	\centering
	\caption{Results: Star Formation Rate in STs and LSBs}
	\label{table:4}
	\begin{tabular}{lcc} % four columns, alignment for each
		\hline
		 &   LSBs & STs \\
		\hline
		\\
		GALEX (FUV)($M_\odot yr^{-1}$) \footnote{median SFR with the "-" and the "+" signs indicating its \\ difference from the first and the third quantiles respectively} & $0.2^{+1.9}_{-0.1}$ & $0.1^{+1.8}_{-0.0}$  \\
		\\
		WISE W3 (NIR) ($M_\odot yr^{-1}$) & $0.3^{+2.1}_{-0.1} $ &$0.3^{+2.2}_{-0.1}$  \\
		\\
		SED\ fitting (MAGPHYS) ($M_\odot yr^{-1}$)&$0.4^{+2.2}_{-0.3} $& $0.2^{+0.9}_{-0.2}$ \\
		\\
		\hline
		
	\end{tabular}}

\end{minipage}
\end{table}

\noindent We present the input magnitudes in ten bands for SED fitting using MAGPHYS for our sample LSBs and STs in Tables A4 and A5 respectively. The modelled star formation properties are presented in Tables A6 and A7 respectively. Figure 2 (Bottom)  shows the histogram of SFR obtained from MAGPHYS. The median of SFR from MAGPHYS is $0.4^{+2.2}_{-0.3} $ $M_{\odot} yr^{-1}$ for LSBs and $0.2^{+0.9}_{-0.2}$ $M_{\odot} yr^{-1}$ for STs, both of which are one to two orders of magnitude less than the SFR for SINGS galaxies (Spitzer Infrared Nearby Galaxies Survey \citep{2008MNRAS.388.1595D}.  Here again we note that the median SFR of LSBs is higher than that of STs.
This can again be attributed to dust-extinction effects predominant in STs. MAGPHYS SED fitting is dominated by optical bands which are prone to dust extinction. \\

\noindent \textbf{Star formation history from MAGPHYS}: The distribution of the random number of bursts after $t_{form}$ is shown in Figure 3 (Left), the respective median values both STs and LSBs being 1, indicating bursts do not play a significant role in modelling the star formation history of LSBs and STs. The distribution of star formation time scale parameter $Gyr^{-1}$ is presented in Figure 3 (Right). The $\gamma$ in $Gyr^{-1}$ is distributed between 0.01 and 0.8. The median values for both STs and LSBs are $\sim$ 0.2 Gyr$^{-1}$, in comparison, $\gamma = 2 Gyr^{-1}$ for an elliptical, $\gamma = 0.25 Gyr^{-1}$ for an early type and $0.5 Gyr^{-1}$  for a late type spiral galaxy \citep{2008MNRAS.388.1595D}.$ \gamma \sim 0 Gyr^{-1}$ implies a constant SFR along the star formation history (SFH) of the galaxy. Among our sample STs and LSBs, most of the galaxies have $ \gamma < 0.5 Gyr^{-1}$ which implies these galaxies have SFHs corresponding to late type or early type spiral galaxies. The minimal number of bursts coupled with the low star formation time scale parameter required in modelling SFH indicates constant SFRs in these galaxies. \\

\noindent \textbf{Signatures of possible dust extinction in STs:}
\noindent As shown in Figure 4 (Top Left), interestingly, most of the galaxies in both the LSB and the ST samples lie in the star forming region with $log(sSFR)(yr^{-1}) > -10.8$ in the $sSFR-M_*$ plane \citep {Salim_2014}. The median of the $log_{10}(sSFR(yr^{-1}))$ distribution is -10.38 and -9.87 for STs and LSBs respectively which shows these are star forming galaxies. Our estimated $log(sSFR)(yr^{-1})$ for STs roughly matches the value -10.56 estimated by \cite{Melnyk_2017}. Further \cite{Karachentseva_2020} used GALEX FUV luminosity to determine the SFR for 181 nearly face-on thin spiral galaxies ($\rm{log}(a/b) < 0.05 $) and found the median $log(sSFR)(yr^{-1})$ -10.40 for Sd galaxies. This complies with the $sSFR$ of our sample STGs and LSBs. This is interesting given the sample of \cite{Karachentseva_2020} consists of galaxies with low dynamical masses compared to our sample STs and LSBs. We further observe that for $M* > 10^9$, at a given stellar mass, the sSFR of an ST is on an average, distinctly less than that of an LSB. Figure 4 [Top Middle] shows the scatter diagram of $log(SFR(M_{\odot} yr^{-1}))$ and $log(M_{*}(M_\odot))$. 
Here also we find a similar trend: at $M* > 10^9$, at a given stellar mass, the SFR of an ST is on an average, distinctly less than that of an LSB.  
Also, we note that most of the STs and the LSBs in the MAGPHYS sample have $M* > 10^9$.

Now dust extinction effects are not well-constrained in our MAGPHYS study due to the non-inclusion of the photometric data beyond 3$\mu m$.  The W3 band was also not included in our MAGPHYS analysis due to the non-availability of significant number of sample galaxies with photometric data in the W3 band as well as the other bands. In any case, the W3 band can trace the hot dust whereas the dust mass is dominated by cold dust. Therefore, it would not be an exaggeration to say that SFR modelled by MAGPHYS are not adequately corrected for dust extinction. This may lead to a possible underestimation of the SFR of the sample galaxies. In fact, the effect is expected to be severe for STs due to their edge-on geometry. Since dust mass scales with the mass of the galaxy, one plausible explanation for the above trend could be dust extinction in STs not adequately accounted for (See \S 4.2 and \S 5).

Figure 4 (Top Right) shows the $NUV-r$ color magnitude diagram of the LSBs and STs in our study with sample used in \cite{Wyder_2009}. They have little UV emission from older stars ($M_{NUV}-M_r>4$). UV luminosity is an indicator of the recent SFR and the $r$-band luminosity traces the total stellar mass, so ($NUV-r$) color is representative of the SFR divided by the stellar mass $M_{star}$(sSFR), and hence the average age of the stars in a galaxy \citep{Salim_2005}. Therefore, at a first glance, the figure may indicate that the STs have an older population of stars compared to LSBs. However, the dust attenuation makes the relation between (NUV-r) color and $SFR/M_{star}$ complicated. Therefore, in conformity with Figures 4 [Top Left] and [Top Middle], the higher values of $NUV – r$ color at a given $r$ may be merely a reflection of higher dust extinction effects in the $NUV$ band in STs as compared to the LSBs
Finally, the Figure 4 on the bottom left shows the distribution of $U-B$ color for LSBs and STs obtained from HyperLeda database. The median $U-B$ color values are 0.28 and -0.06 for STs and LSBs. The Figure 4 on right shows the distribution of $B-V$ color, the median $B-V$ color value is 0.9 and 0.6 for STs and LSBs. This may again be an indication of either a predominantly older stellar population in STs compared to LSBS, or alternatively, higher dust extinction effects in STs compared to LSBs. See \cite{1999AJ....118.2751M} and \cite{Matthews_2001}.)  \\

\noindent \textbf{Comparison with HSBs:} For the sake of comparison, we carry out the star formation analysis for two sets of high surface brightness galaxies: face-on and edge-on. We choose 212 HSBs (with an angle of inclination $i$ $<$ 45$^0$ and $B$-band surface brightness $\mu_b > 23 \rm{mag arcsec}^{-2}$) from the UGC, which constitute our face-on HSB set; 69 HSBs
(with an angle of inclination $i$ $\sim$ 90$^0$ and $B$-band surface brightness $\mu_b > 23   {\rm{mag\ arcsec}}^{-2}$) from the RFGC constitute our edge-on HSB set. From SED fitting using MAGPHYS, we find that the median SFR for the for face-on HSBs and edge-on HSBs are 3.4 and 1.6 $M_{\odot} yr^{-1}$ respectively. We recall that the corresponding values for the face-on LSBs and edge-on LSBs (STs) are 0.4 and 0.2 $M_{\odot} yr^{-1}$ respectively. Thus, the SFRs of LSBs and STs are, on an average, several times lower those of normal (HSB) face-on and edge-on galaxies.  Further, the median specific star formation rate $log_{10}sSFR$ of our face-on HSBs and edge-on HSBs are -9.1 $M_{\odot} yr^{-1}$ and -9.6 $M_{\odot} yr^{-1}$ respectively, as opposed to -9.9 and -10.4 per year respectively for our LSBs and STs. Therefore, we conclude, that LSBs and STs are not part of the population of normal HSB galaxies, as far as SFR is concerned.

%\end{itemize}

\subsection{Are the STs LSBs seen edge-on?:} 
\cite{2011Maclachlan} used radiative transfer models to derive the dust disc properties of 3 edge-on LSBs ( STs ): UGC7321, IC2233 and NGC 4244 in $B$ band, and found that the dust discs had smaller scale lengths relative to their stellar scale heights, which rendered them difficult to be detected in observations. One can easily estimate the SFR for the above 3 STs, also part of our sample, when seen face-on. 
Using the de-projected central surface brightness (uncorrected for edge-on dust extinction) (Table 1 of \cite{2011Maclachlan}), we apply the dust extinction correction corresponding to edge-on (Table 4 of \cite{2011Maclachlan}) to obtain the de-projected central surface brightness (corrected for dust extinction). Next, we add the dust extinction due to face-on opacity (Table 3 of \cite{2011Maclachlan}) on the de-projected central surface brightness (corrected for dust extinction) to obtain de-projected central surface brightness (uncorrected for face-on dust extinction), which would have been the observed surface brightness if the ST were seen face-on. 
The central surface brightness $S$  is related to apparent magnitude $m$, the extinction co-efficient $A_{\lambda}$  and aperture area $A$ as 
$S=m+ A_\lambda + 2.5\rm{log}(A)$;
where $A_{\lambda}$ is related to opacity $\tau_{\lambda}$ as $A_{\lambda} = 1.086 \tau_{\lambda}$. 

We determine the central surface density and hence the mass of the corresponding stellar disc using the above relations for the following two cases: (1) Projected central surface brightness (uncorrected for any dust extinction) of the sample ST (2) De-projected central surface brightness (with face-on dust extinction). The first case corresponds to the mass of an ST determined from its observed surface brightness, uncorrected for dust extinction effects. The second case corresponds to the same mass determined for the same ST but when seen face-on.

Finally, using the $M_*-SFR$ correlation given in \cite{2008Dave} as obtained from cosmological hydrodynamical simulations, we obtain the SFRs corresponding to the above masses. We find that the SFR for an ST when seen face-on is higher than that seen edge-on by 0.2 - 1.3 $M_{\odot} yr^{-1}$, the median change in SFR being 0.3 $M_{\odot} yr^{-1}$ . The results of the calculations are tabulated in Table 5. Interestingly, this roughly matches the difference in the SFR of an ST and an LSB at a given mass (Figure 4 [Top, Middle]). We note, our results indicate that the surface brightness of our STs when seen face-on is not $>$ 23 $\rm{mag {arcsec}^{-2}}$ as per the definition of LSBs in the strict sense of the term
but on the borderline (See Tables A1 and A2).
However, our calculation is just an illustrative exercise to roughly estimate the dust extinction effects in STs. This may possibly explain the systematically lower SFRs in STs as compared to LSBs at a given mass.\emph{This possibly indicates that an ST may very well be an LSB as seen edge-on, thereby explaining the systematically lower SFR(sSFR) values of STs compared to LSBs at higher masses}.

\begin{figure*}
\begin{center}
\begin{tabular}{cccc}
\resizebox{65mm}{!}{\includegraphics{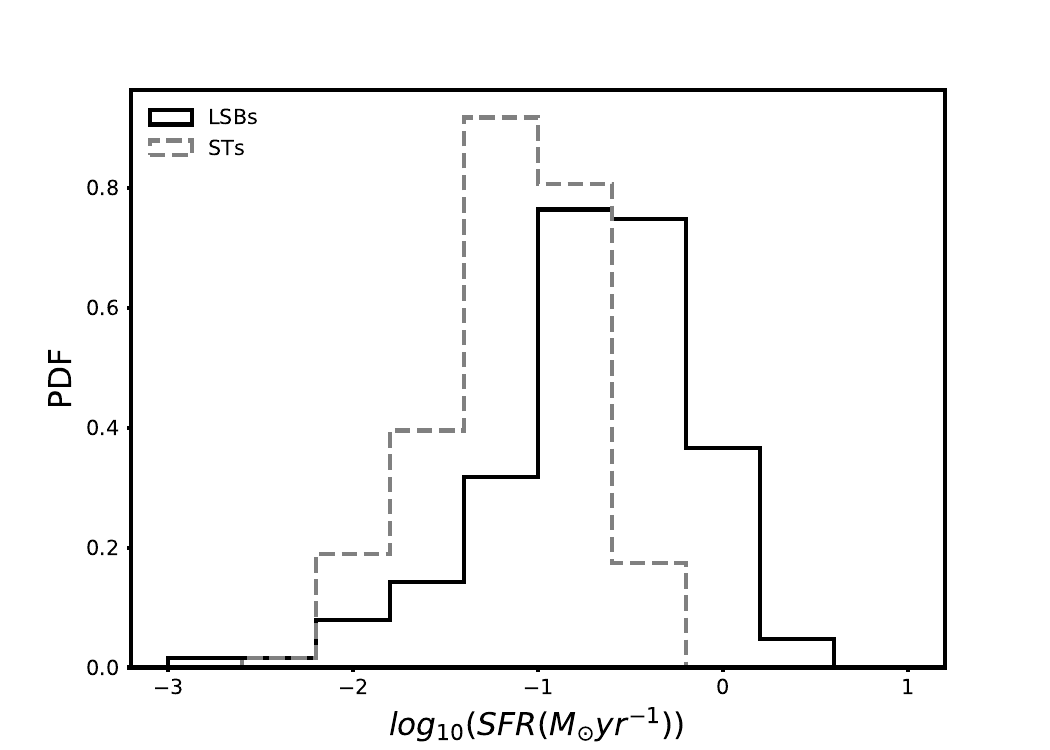}} 
\resizebox{65mm}{!}{\includegraphics{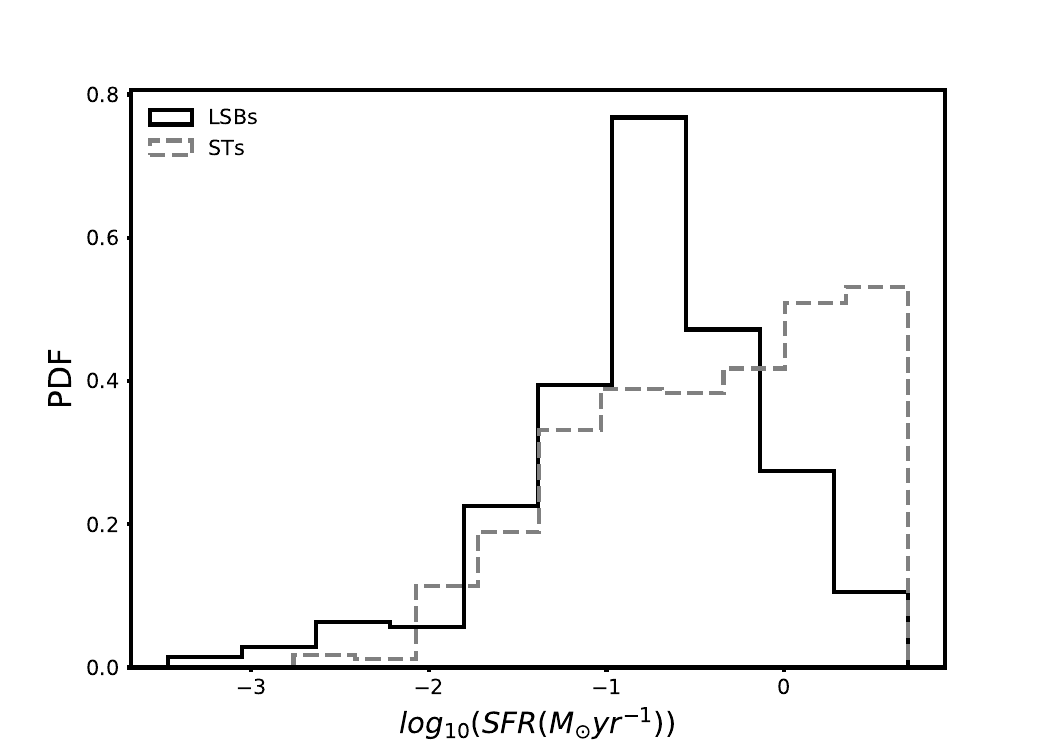}} \\
\resizebox{65mm}{!}{\includegraphics{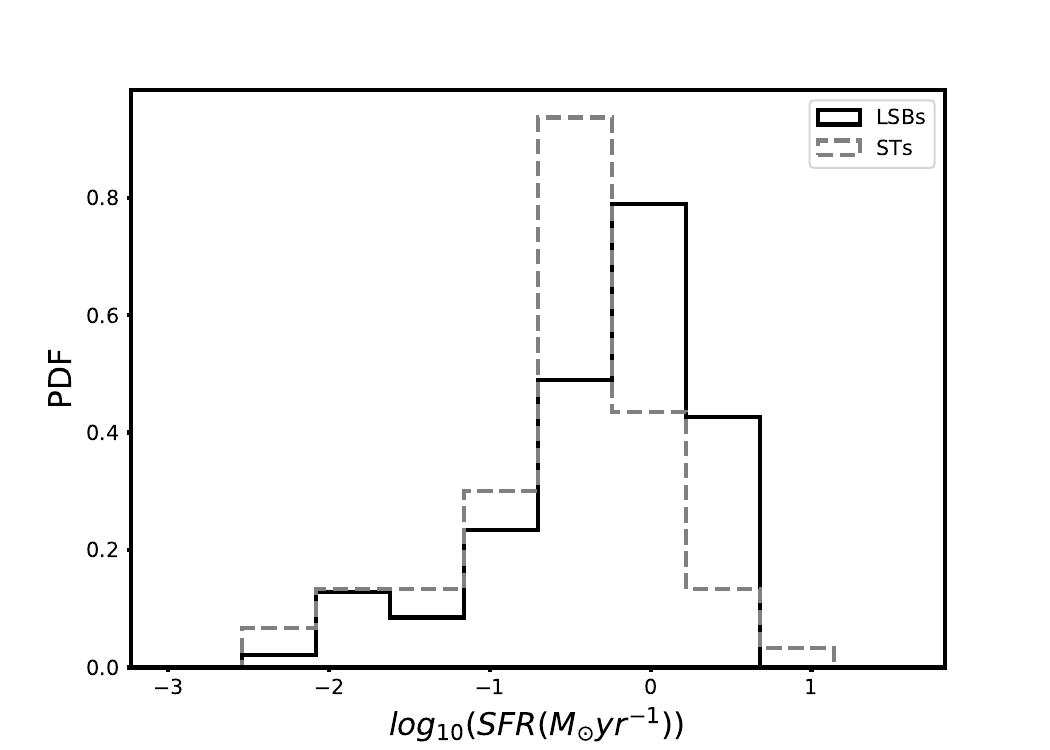}}\\
\end{tabular}
\end{center}
\caption{The above figures show the distributions of SFR in STs and LSBs as estimated from different tracers:
Top Left: GALEX FUV, Top Right: WISE NIR, Bottom: SED Fitting (MAGPHYS) }
\label{fig:1}
\end{figure*}

\begin{figure*}
\begin{center}
\begin{tabular}{cc}
\resizebox{65mm}{!}{\includegraphics{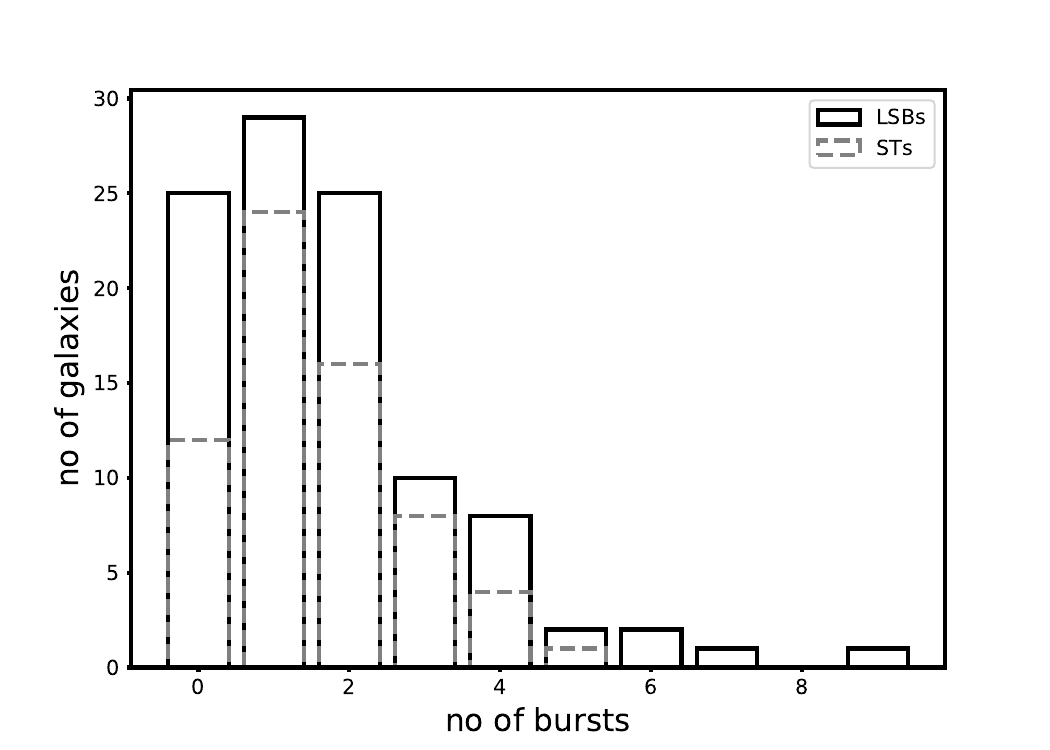}}
\resizebox{65mm}{!}{\includegraphics{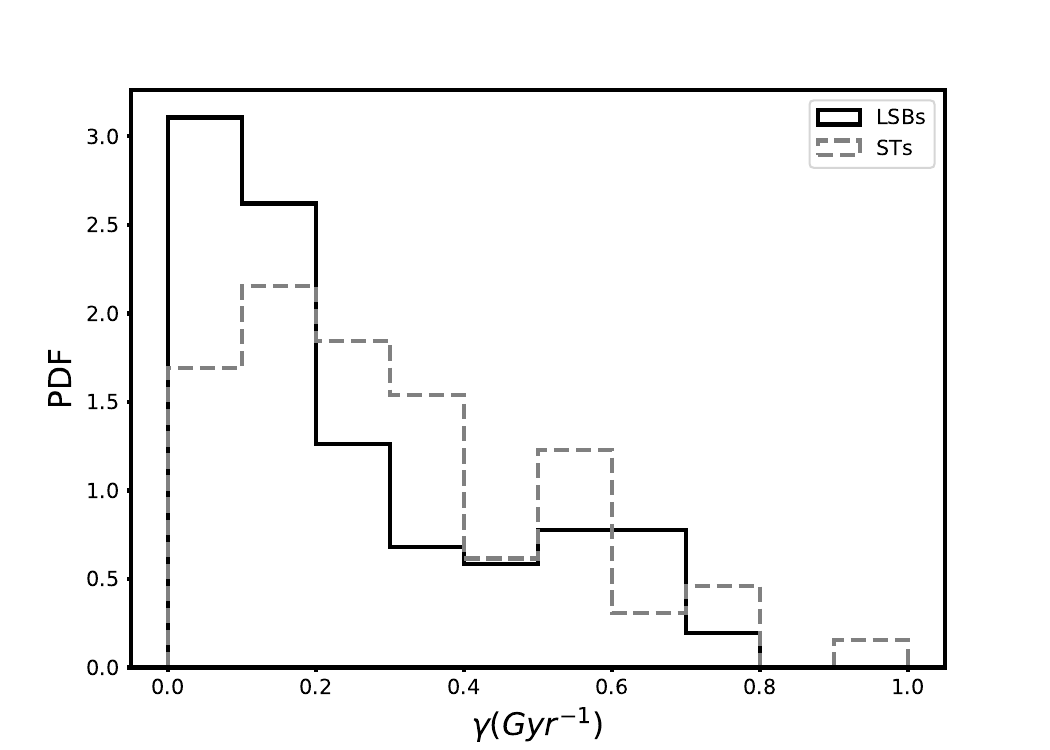}}
\end{tabular}
\end{center}
\caption{ [Left]: Distribution of number of bursts after $t_{form}$ and [Right]: Distribution of timescale of star formation $\gamma$ in $Gyr^{-1}$ as modelled by SED fitting using MAGPHYS in our sample STs and LSBs.}
\label{fig:5}
\end{figure*}

\begin{table}
\centering
\begin{minipage}{150mm}
\caption{Results: SFR of STs after dust extinction correction}
\label{table:5}
\begin{tabular}{lcccccc} % four columns, alignment for each
\hline
Galaxy & $\mu^1$ \footnote{Edge-on Surface Brightness (with dust extinction) ($\rm{mag} / \rm{arcsec}^{-2}$)} & $\mu^2$ \footnote{Face-on Surface Brightness (with dust extinction) ($\rm{mag} / \rm{arcsec}^{-2}$)} & $\tau^1$ \footnote{Face-on opacity}& $\tau^2$ \footnote{Edge-on opacity} & $SFR^1$ \footnote{Star formation rate (edge-on) in $M_{\odot} yr^{-1}$}& $SFR^2$ \footnote{Star formation rate (face-on) in $M_{\odot} yr^{-1}$}\\
\hline
UGC 7321 & 23.45 & 21.51 & 0.072 & 1.99 & 0.1 & 0.5 \\
    IC 2233 & 22.73 & 21.88 & 0.034 & 0.69 & 0.3 & 0.5 \\
    NGC 4244 & 24.51 & 22.01& 0.106 & 1.57 & 0.1 & 0.4 \\
\hline
\end{tabular}
\end{minipage}
\end{table}

\begin{figure*}
\begin{center}
\begin{tabular}{cccc}
\resizebox{60mm}{!}{\includegraphics{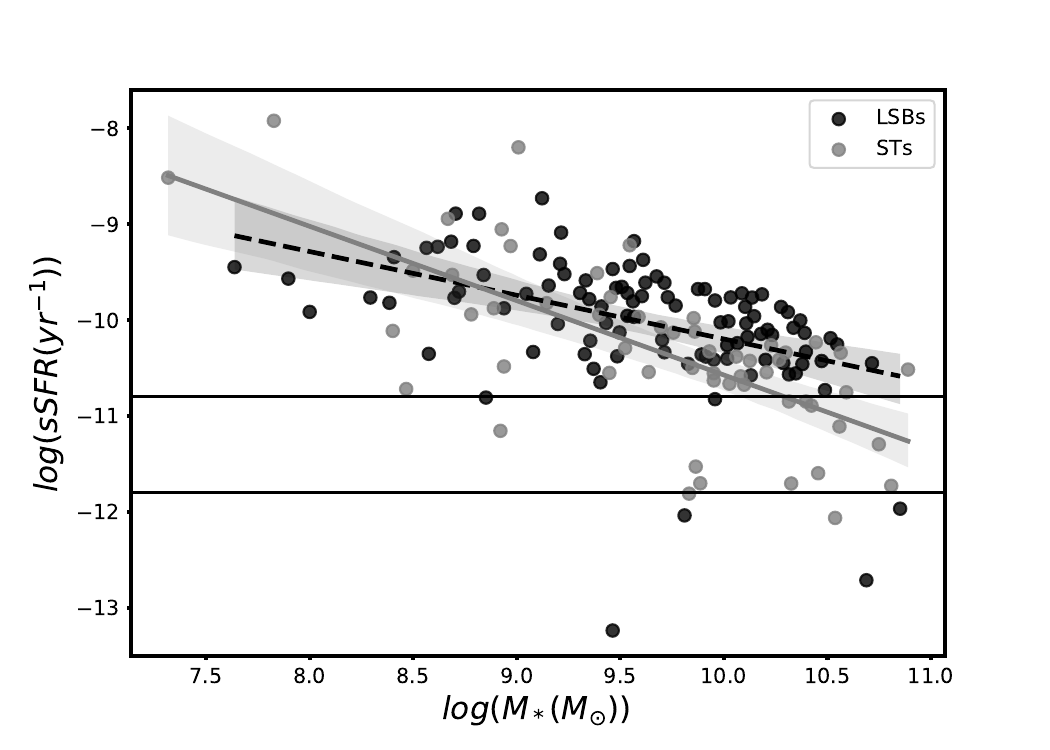}}
\resizebox{60mm}{!}{\includegraphics{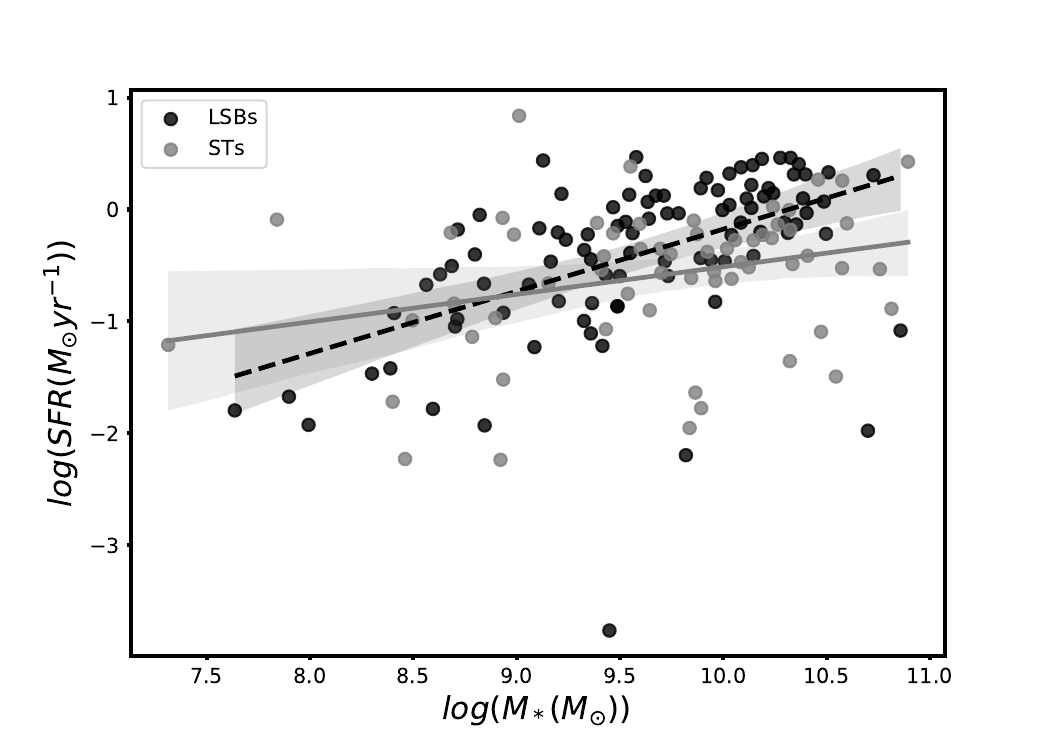}}
\resizebox{60mm}{!}{\includegraphics{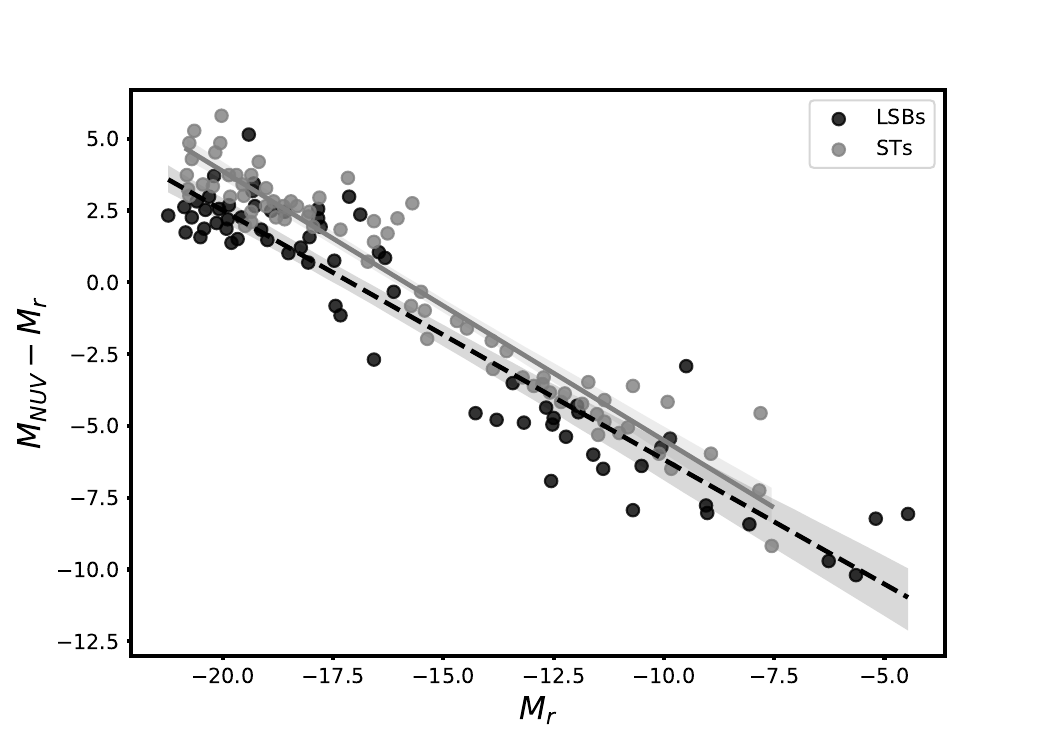}}
\\
\resizebox{60mm}{!}{\includegraphics{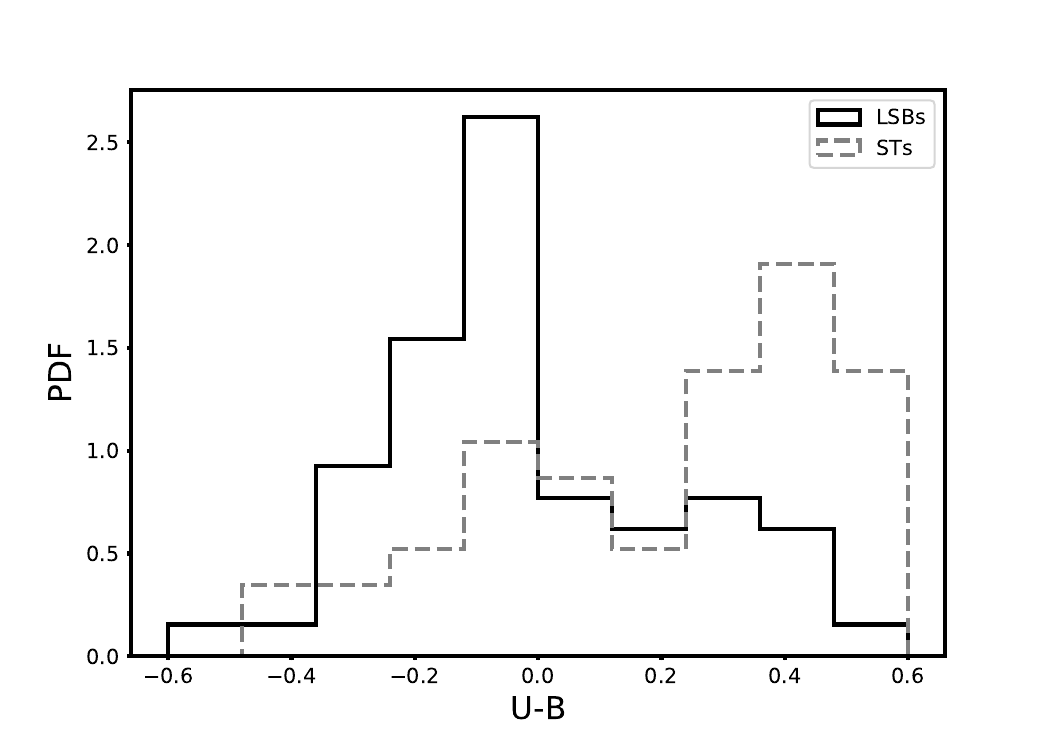}}
\resizebox{60mm}{!}{\includegraphics{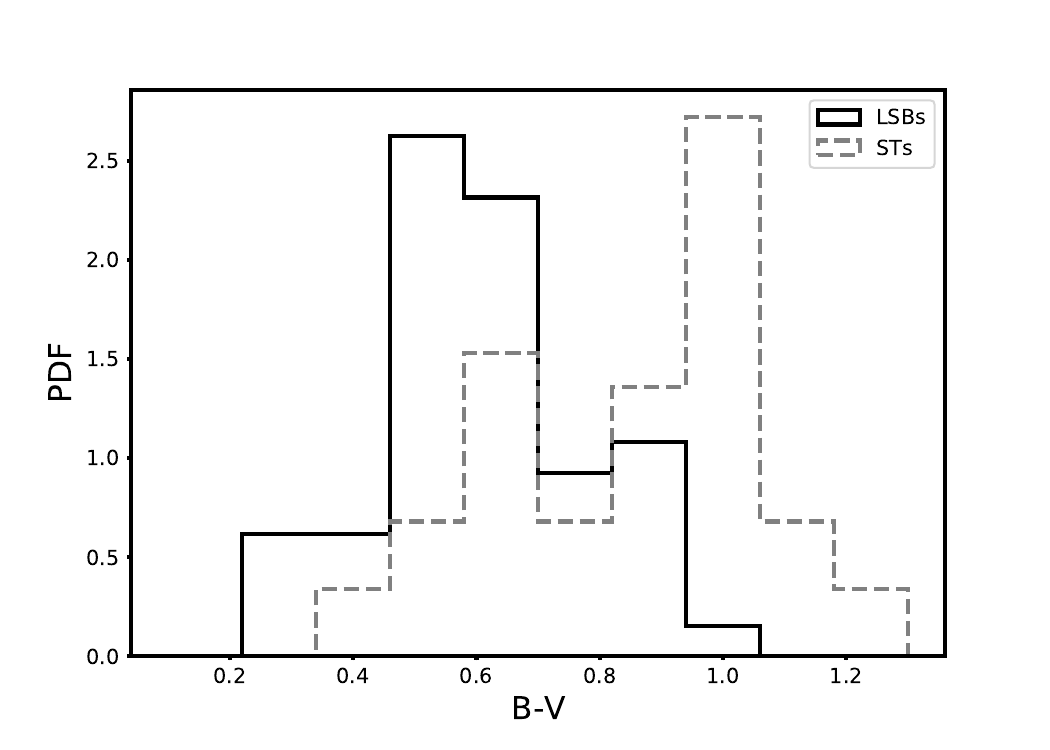}}
\end{tabular}
\end{center}
\caption{[Top Left:] Scatter plot of  $\rm{log}(sSFR)(yr^{-1})$ and $\rm{log}(M_*)$ for STs and LSBs estimated from SED fitting using MAGPHYS for our sample STs and LSBs. The dashed and the solid lines indicate the linear fits to the ST and the LSB data respectively.[Top Middle:] Scatter plot of  $\rm{\mathbf{log(SFR)(M_\odot yr^{-1}})}$ and $\rm{log}(M_*)$ for STs and LSBs estimated from SED fitting using MAGPHYS for our sample STs and LSBs. The dashed and the solid lines indicate the linear fits to the ST and the LSB data respectively.[Top Right:] Color (NUV-r) magnitude (r) diagram for our ST and LSB samples used for SED fitting study. The dashed and the solid lines indicate the linear fits to the ST and the LSB data respectively.[Bottom Left:]  Distribution of U-B color for  STs and LSBs from HyperLeda database. 
[Bottom Right:]  Distribution of B-V color for  STs and LSBs from HyperLeda database.  }
\label{fig:2}
\end{figure*}

\section{Discussion}
\begin{itemize}

    \item \underline{Aperture photometry:} In this study, we have used multi-band photometry from the corresponding catalogues to determine the SFR. However, this may lead to erroneous results when applied to nearby galaxies. This is true in particular for galaxies like the STs as these galaxies are not well-fitted by elliptical apertures (\cite{Melnyk_2017}). To check for this, we take a small subset of our GALEX FUV ST and LSB sample, and determine the magnitudes using aperture photometry. Our results indicate that the magnitudes determined by the two different techniques are not significantly different. \\

    \item \underline{WISE W3 vs WISE W4:} In this paper, we have used WISE W3 photometry to trace SFR in NIR, although WISE W3 might include a significant contribution from silicates, either in absorption or emission, it might miss a fraction of the dust heating. However, WISE W4 photometry, despite its lower spatial resolution, could recover that additional heating. But, it could, on the other hand, lead to an over-estimation of the SFR due to AGN heating \cite{Cluver_2017}.  To test for this, we calculate the SFR using WISE W4 luminosity for a subset of our WISE W3 sample of STs and LSBs, using Equation 6 from \cite{Cluver_2017} for SFR luminosity estimation. Interestingly, the median SFRs for both LSBs and STs as determined from the W4 band hardly differ from that estimated from the W3 band, thus confirming the reliability of the W3 band as an SFR tracer. \\

\item  \underline{SED fitting including WISE W3 band:} In the current study, we have not included WISE W3 band in the SED fitting using MAGPHYS. This was because the subset of our LSB and ST sample with data in the current ten photometric bands plus WISE band was very small: 45 LSBs and 39 STs. However, to check for the effect of W3, we determine the SFR using SED fitting with the current ten photometric bands plus W3, and found that the median SFR for both the STs and the LSBs increase slightly to $\sim$ 0.7 $M_{\odot} yr^{-1}$. 

\end{itemize}
\section{Conclusions}

Superthin galaxies (STs) are edge-on disc galaxies with strikingly high planar-to-vertical axes ratios of $\sim 10 - 20$ with no bulge component, and central surface brightness in $B$-band $>$ 23 magarcsec$^{-2}$ comparable to low surface brightness galaxies (LSBs). Although STs and LSBs have almost comparable median dynamical mass, stellar mass and atomic hydrogen HI mass, it is tricky to conclude if they constitute one and the same galaxy population. This is due to the ambiguity in modelling the surface densities and vertical scale heights due to the edge-on and face-on orientations of the STs and the LSBs respectively. 
%However, SFR can serve as a useful probe of the vertical scale height or thickness of LSBs as compared to STs as star formation is driven by the disc dynamical instability against local, axi-symmetric perturbations. According to the disc dynamical stability models of 2-component galactic discs, a disc with a larger vertical scale height \citep{1996MNRAS.278..209J} and  vertical velocity dispersion  \citep{Romeo_2011} results in a stabler disc, resulting in a lower SFR. In other words, a systematic study of SFR could be used to study the disc scale heights in face on galaxies like the LSBs which are not amenable to direct observational estimation.
We systematically study star formation in a sample of LSBs and STs using multi-wavelength data and stellar population synthesis models to compare and contrast the star formation properties of LSBs and STs. The SFRs of LSBs and STs as determined by different observational tracers are as follows: GALEX FUV: $0.2^{+1.9}_{-0.1}$ $M_\odot yr^{-1}$ \& $0.1^{+1.8}_{-0.0}$ $M_\odot yr^{-1}$, WISE NIR: $0.3^{+2.1}_{-0.1} $ $M_\odot yr^{-1}$ \& $0.3^{+2.2}_{-0.1}$ $M_\odot yr^{-1}$, MAGPHYS (SED fitting of photometric data in ten bands including FUV, NUV of GALEX, u,g,r,i,z of SDSS \& J, H, Ks of 2MASS)]: $0.4^{+2.2}_{-0.3} $ $M_\odot yr^{-1}$ \& $0.2^{+0.9}_{-0.2}$ $M_\odot yr^{-1}$ ordinary galaxies. Further, from SED fitting using MAGPHYS, we find that the median SFR for the for a sample of face-on HSBs and edge-on HSBs are 3.4 and 1.6 $M_{\odot} yr^{-1}$ respectively. Therefore, although the SFR of STs and LSBs are comparable, they are several times lower than that of than ordinary HSBs. Also, as is indicated by the median values of number of bursts after $t_{\rm{form}} = 1$ and an exponential star formation time scale parameter $\gamma$ = 0.2 ${\rm{Gyr}}^{-1}$,  the SFR remains fairly constant over time, for both STs and LSBs. Interestingly, in spite of having low SFR compared to ordinary star-forming galaxies, both STs and LSBs populate the star forming blue cloud region in the specific star formation (sSFR) - stellar mass (M*) plane. Further, at higher masses, STs have a lower sSFR as well as SFR compared to LSBs.  Our calculations indicate that the SFR of a ST may increase by 0.3 $M_{\odot}\rm{yr}^{-1}$ if we apply inclination correction and account for appropriate opacity corrections, and therefore can explain the mismatch between the SFRS of LSBs and STs.

\section{Data availability}

We obtained the data  from the catalog  \cite{refId0}, for LSBs and \cite{karachentsev2003revised}, for STs using Vizier catalogue access tool. A python package ASTROQUERY was used to obtain the photometry data in all bands. Distances to the galaxies were obtained using the NASA/IPAC Extragalactic Database (NED). Physical properties of the galaxies were downloaded from HyperLeda\footnote{http://leda.univ-lyon1.fr/} database.

\section{Acknowledgements}
This research has made use of the VizieR catalogue access tool, CDS, Strasbourg, France (DOI : 10.26093/cds/vizier). The original description of the VizieR service was published in 2000, A\&AS 143, 23. This research we use the NASA/IPAC Extragalactic Database (NED), which is operated by the Jet Propulsion Laboratory, California Institute of Technology,under contract with the National Aeronautics and Space Administration.
In this study we use MAGPHYS - (Multi-wavelength Analysis of Galaxy Physical Properties) a open public software to fit the observed spectral energy distributions of galaxies \citep{2008MNRAS.388.1595D}.
We use the publicly available software Astropy a community-developed core Python package for Astronomy \citep{astropy:2018} for handling FITS files.  ASTROQUERY \citep{Ginsburg_2019} an - ASTROPY-affiliated PYTHON package  for  accessing  remotely  hosted  astronomical  data, was used for downloading the catalog from vizier. The PHOTUTILS\citep{larry_bradley_2020_4044744} was used to aperture photometry for few of the galaxies.\\

The authors would also like to thank Dr. Peter Kamphuius, Dr.Sudanshu Barway, Dr.Koshy George, Dr.Omkar Bait and Prof. Nissim Kanekar for useful suggestions and discussion.

\small{\bibliographystyle{mnras}} 
\bibliography{references.bib}
\onecolumn
\newpage
\appendix

\begin{table}
\textbf{APPENDIX}
\renewcommand\thetable{}
    \centering
	\label{Appendix table:1}
    \caption{\textbf{A1} Physical properties LSBs }
	\begin{tabular}{lccccccc}
	\hline
	LSB \footnote{Name of the galaxy}& $M_B$ \footnote{B-band apparent magnitude} & Distance\footnote{Distance to galaxy in Mpc} & Vrot\footnote{asymptotic rotational velocity in km/s} & a/b \footnote{log(a/b) major to minor axes ration in 0.1 dec unit}& $\mu_B$ \footnote{Surface brightness in B band in $mag arcsec^{-1}$}&Diameter \footnote{Diamter of the galaxy in kpc} \\
	\hline
$	PGC	1241	$&$	14.56	$&$	58.03	$&$	67.22	$&$	1.31	$&$	23.09	$&$	31.3	$	\\
$	PGC	1288	$&$	15.84	$&$	66.83	$&$	65.13	$&$	1.02	$&$	22.8	$&$	29.0	$	\\
$	PGC	1387	$&$	15	$&$	78.99	$&$	68.13	$&$	1.43	$&$	23.08	$&$	27.0	$	\\
$	PGC	6574	$&$	14.59	$&$	82.75	$&$	65.22	$&$	1	$&$	22.81	$&$	29.7	$	\\
$	PGC	6833	$&$	15.09	$&$	74.27	$&$	69.13	$&$	1.22	$&$	23.17	$&$	28.5	$	\\
$	PGC	7399	$&$	14.11	$&$	55.38	$&$	60.77	$&$	1.11	$&$	23.01	$&$	30.7	$	\\
$	PGC	12889	$&$	14.81	$&$	84.43	$&$	60.83	$&$	1.17	$&$	22.98	$&$	25.5	$	\\
$	PGC	13190	$&$	15.62	$&$	70.55	$&$	61.34	$&$	1.37	$&$	23.21	$&$	29.5	$	\\
$	PGC	961488	$&$	15.78	$&$	63.62	$&$	66.56	$&$	1.4	$&$	22.9	$&$	26.4	$	\\
$	PGC	13889	$&$	14.7	$&$	67.54	$&$	64.9	$&$	1.3	$&$	23.05	$&$	25.7	$	\\
$	PGC	15292	$&$	15.81	$&$	83.63	$&$	68.13	$&$	1.05	$&$	22.83	$&$	28.5	$	\\
$	PGC	15858	$&$	15.09	$&$	67.38	$&$	60.31	$&$	1.04	$&$	23.15	$&$	26.6	$	\\
$	PGC	16702	$&$	14.67	$&$	56.87	$&$	66.44	$&$	1.05	$&$	22.87	$&$	29.1	$	\\
$	PGC	17323	$&$	15.43	$&$	62.01	$&$	60.03	$&$	1.42	$&$	23.02	$&$	29.7	$	\\
$	PGC	17402	$&$	15.56	$&$	76.88	$&$	69.59	$&$	1.14	$&$	23.01	$&$	31.2	$	\\
$	PGC	18099	$&$	14.07	$&$	57.87	$&$	61.36	$&$	1.36	$&$	23.16	$&$	31.9	$	\\
$	PGC	18259	$&$	15.72	$&$	79.1	$&$	64.71	$&$	1.43	$&$	22.89	$&$	25.1	$	\\
$	PGC	18315	$&$	14.51	$&$	57.24	$&$	68.47	$&$	1.06	$&$	22.97	$&$	25.1	$	\\
$	PGC	18377	$&$	15.13	$&$	78.53	$&$	65.84	$&$	1.39	$&$	22.86	$&$	29.9	$	\\
$	PGC	19674	$&$	14.03	$&$	80.08	$&$	62.48	$&$	1.33	$&$	22.87	$&$	29.1	$	\\
$	PGC	20700	$&$	15.84	$&$	61.83	$&$	62.7	$&$	1.41	$&$	23.21	$&$	26.4	$	\\
$	PGC	22678	$&$	14.64	$&$	63.75	$&$	61.04	$&$	1.22	$&$	23.29	$&$	30.6	$	\\
$	PGC	23146	$&$	15.71	$&$	70.59	$&$	62.04	$&$	1.31	$&$	22.81	$&$	26.2	$	\\
$	PGC	23352	$&$	15.67	$&$	80.14	$&$	62.77	$&$	1.28	$&$	23.07	$&$	31.6	$	\\
$	PGC	24328	$&$	14.6	$&$	74.2	$&$	64.01	$&$	1.16	$&$	22.83	$&$	25.1	$	\\
$	PGC	25281	$&$	14.93	$&$	65.46	$&$	69.91	$&$	1.2	$&$	22.98	$&$	30.9	$	\\
$	PGC	25406	$&$	14.46	$&$	71.38	$&$	65.09	$&$	1.07	$&$	23	$&$	30.9	$	\\
$	PGC	25472	$&$	15.64	$&$	61.05	$&$	64.43	$&$	1.13	$&$	23	$&$	25.6	$	\\
$	PGC	1841518	$&$	15.08	$&$	67.79	$&$	60.06	$&$	1.15	$&$	23.17	$&$	29.1	$	\\
$	PGC	26297	$&$	15.34	$&$	71.68	$&$	69.89	$&$	1.24	$&$	23.12	$&$	27.2	$	\\
$	PGC	26752	$&$	14.31	$&$	76.59	$&$	61.22	$&$	1.44	$&$	23.19	$&$	27.2	$	\\
$	PGC	26852	$&$	15.51	$&$	61.21	$&$	60.16	$&$	1.41	$&$	23.1	$&$	28.5	$	\\
$	PGC	27747	$&$	15.42	$&$	72.55	$&$	64.34	$&$	1.43	$&$	23.22	$&$	28.9	$	\\
$	PGC	28830	$&$	15.59	$&$	70.26	$&$	69.51	$&$	1.31	$&$	22.92	$&$	28.2	$	\\
$	PGC	29865	$&$	14.99	$&$	74.73	$&$	61.54	$&$	1.33	$&$	23.1	$&$	26.6	$	\\
$	PGC	29924	$&$	15.51	$&$	57.22	$&$	68.18	$&$	1.28	$&$	23.28	$&$	26.5	$	\\
$	PGC	1041741	$&$	14.31	$&$	80.33	$&$	67.44	$&$	1.09	$&$	23.06	$&$	30.7	$	\\
$	PGC	30818	$&$	15.75	$&$	81.33	$&$	69.79	$&$	1.31	$&$	22.81	$&$	27.4	$	\\
$	PGC	31122	$&$	15.68	$&$	83.79	$&$	66.7	$&$	1.37	$&$	23.25	$&$	29.6	$	\\
$	PGC	32519	$&$	15.6	$&$	78.92	$&$	64.51	$&$	1.12	$&$	23.21	$&$	28.4	$	\\
$	PGC	32946	$&$	15.03	$&$	64.79	$&$	60.37	$&$	1.33	$&$	22.99	$&$	26.3	$	\\
$	PGC	34556	$&$	14.61	$&$	83.98	$&$	62.94	$&$	1.45	$&$	23.16	$&$	28.4	$	\\
$	PGC	34681	$&$	14.84	$&$	69.53	$&$	65.62	$&$	1.37	$&$	22.87	$&$	27.0	$	\\
$	PGC	35900	$&$	14.98	$&$	55.18	$&$	64.35	$&$	1.12	$&$	23.24	$&$	28.6	$	\\
$	PGC	36008	$&$	15.07	$&$	83.99	$&$	66.13	$&$	1.05	$&$	22.96	$&$	28.3	$	\\
$	PGC	36079	$&$	15.09	$&$	80.17	$&$	60.64	$&$	1.35	$&$	23.01	$&$	31.5	$	\\
$	PGC	36929	$&$	14.45	$&$	71.1	$&$	67.13	$&$	1.1	$&$	23.18	$&$	25.9	$	\\
$	PGC	37308	$&$	15.3	$&$	76.39	$&$	63.73	$&$	1.39	$&$	23.19	$&$	28.7	$	\\
$	PGC	37629	$&$	14.49	$&$	72.87	$&$	64.66	$&$	1.25	$&$	22.97	$&$	31.3	$	\\
$	PGC	37692	$&$	15.63	$&$	67.58	$&$	67.57	$&$	1.22	$&$	22.94	$&$	25.4	$	\\
$	PGC	37832	$&$	14.19	$&$	83.73	$&$	67.75	$&$	1.04	$&$	22.81	$&$	30.5	$	\\
$	PGC	42999	$&$	15.74	$&$	61.21	$&$	68.61	$&$	1.08	$&$	22.97	$&$	28.1	$	\\
$	PGC	43359	$&$	14.61	$&$	80.27	$&$	66.6	$&$	1.5	$&$	23.15	$&$	29.6	$	\\
$	PGC	43385	$&$	15.79	$&$	74.51	$&$	64.51	$&$	1.14	$&$	23.15	$&$	31.0	$	\\
$	PGC	43458	$&$	15.3	$&$	69.97	$&$	64.18	$&$	1.02	$&$	23.08	$&$	29.8	$	\\
$	PGC	44014	$&$	15.22	$&$	69.56	$&$	67.72	$&$	1.23	$&$	22.98	$&$	27.2	$	\\
$	PGC	45644	$&$	15.62	$&$	84.66	$&$	67.71	$&$	1.43	$&$	22.82	$&$	31.7	$	\\
$	PGC	45684	$&$	15.21	$&$	83.48	$&$	68.1	$&$	1.31	$&$	22.84	$&$	25.7	$	\\
$	PGC	45992	$&$	15.29	$&$	74.42	$&$	67.09	$&$	1.48	$&$	23.1	$&$	26.9	$	\\
$	PGC	46589	$&$	14.2	$&$	62.58	$&$	60.62	$&$	1.09	$&$	22.86	$&$	28.5	$	\\
$	PGC	47231	$&$	15.53	$&$	75.27	$&$	66.2	$&$	1.28	$&$	22.87	$&$	25.6	$	\\
$	PGC	47255	$&$	15.73	$&$	72.04	$&$	61.31	$&$	1.07	$&$	22.88	$&$	25.1	$	\\
$	PGC	47368	$&$	14.61	$&$	59.45	$&$	68.26	$&$	1.31	$&$	22.97	$&$	29.0	$	\\
\end{tabular}
\end{table}

\begin{table}
\renewcommand\thetable{}
\centering
    \caption{\textbf{A2} General Properties: STs}
    \label{Appendix : Table 2} 
    \begin{tabular}{lccccccc} 
    \hline
	ST \footnote{Name of the galaxy}& $M_B$ \footnote{B-band apparent magnitude} & Distance\footnote{Distance to galaxy in Mpc} & Vrot\footnote{asymptotic rotational velocity in km/s} & a/b \footnote{log(a/b) major to minor axes ratio in 0.1 dec unit \label{tab:table3}\\ }& $\mu_B$ \footnote{Surface brightness in B band in $mag arcsec^{-1}$}&Diameter \footnote{Diamter of the galaxy in kpc} \\
	\hline
$	RFGC	6	$	&	$	15.31	$	&	$	79.07	$	&	$	71.64$	&	$		1.31	$	&	$	23.29	$	&	$	23.0	$	\\
$	RFGC	34	$	&	$	16.22	$	&	$	66.77	$	&	$	66.46$	&	$		1.11	$	&	$	23.26	$	&	$	28.4	$	\\
$	RFGC	45	$	&	$	16.8	$	&	$	77.89	$	&	$	68.90$	&	$		1.03	$	&	$	23.11	$	&	$	28.7	$	\\
$	RFGC	51	$	&	$	16.54	$	&	$	74.48	$	&	$	64.15$	&	$		1.07	$	&	$	22.94	$	&	$	28.9	$	\\
$	RFGC	73	$	&	$	15.12	$	&	$	94.98	$	&	$	63.03$	&	$		1.23	$	&	$	22.92	$	&	$	32.4	$	\\
$	RFGC	81	$	&	$	14.61	$	&	$	67.58	$	&	$	71.85$	&	$		1.32	$	&	$	22.95	$	&	$	28.4	$	\\
$	RFGC	99	$	&	$	16.91	$	&	$	90.55	$	&	$	66.22$	&	$		1.06	$	&	$	23.15	$	&	$	23.2	$	\\
$	RFGC	112	$	&	$	14.25	$	&	$	73.73	$	&	$	63.86$	&	$		1.43	$	&	$	23.29	$	&	$	25.2	$	\\
$	RFGC	113	$	&	$	14.6	$	&	$	66.53	$	&	$	69.62$	&	$		1.31	$	&	$	23.12	$	&	$	24.9	$	\\
$	RFGC	122	$	&	$	14.97	$	&	$	83.79	$	&	$	69.35$	&	$		1.29	$	&	$	23.18	$	&	$	23.4	$	\\
$	RFGC	123	$	&	$	16.94	$	&	$	73.17	$	&	$	71.81$	&	$		1.34	$	&	$	23.28	$	&	$	34.0	$	\\
$	RFGC	139	$	&	$	16.37	$	&	$	72.22	$	&	$	67.61$	&	$		1.34	$	&	$	23.18	$	&	$	32.6	$	\\
$	RFGC	155	$	&	$	14.52	$	&	$	69.59	$	&	$	68.91$	&	$		1.11	$	&	$	23.02	$	&	$	24.9	$	\\
$	RFGC	161	$	&	$	15.56	$	&	$	82.56	$	&	$	64.72$	&	$		1.07	$	&	$	22.92	$	&	$	24.2	$	\\
$	RFGC	164	$	&	$	16.15	$	&	$	69.93	$	&	$	64.85$	&	$		1.46	$	&	$	22.97	$	&	$	34.2	$	\\
$	RFGC	168	$	&	$	15.19	$	&	$	79.54	$	&	$	68.89$	&	$		1.38	$	&	$	23.28	$	&	$	29.9	$	\\
$	RFGC	176	$	&	$	15.38	$	&	$	94.26	$	&	$	65.30$	&	$		1.39	$	&	$	23.38	$	&	$	33.4	$	\\
$	RFGC	184	$	&	$	14.38	$	&	$	89.52	$	&	$	65.54$	&	$		1.49	$	&	$	23.26	$	&	$	30.2	$	\\
$	RFGC	191	$	&	$	16.91	$	&	$	82.03	$	&	$	68.04$	&	$		1.31	$	&	$	23.12	$	&	$	23.6	$	\\
$	RFGC	207	$	&	$	15.12	$	&	$	66.77	$	&	$	69.36$	&	$		1.32	$	&	$	23.15	$	&	$	31.0	$	\\
$	RFGC	213	$	&	$	14.92	$	&	$	67.16	$	&	$	70.00$	&	$		1.31	$	&	$	23.27	$	&	$	23.1	$	\\
$	RFGC	261	$	&	$	16.91	$	&	$	86.34	$	&	$	65.31$	&	$		1.15	$	&	$	23.2	$	&	$	24.5	$	\\
$	RFGC	278	$	&	$	14.87	$	&	$	69.73	$	&	$	69.67$	&	$		1.12	$	&	$	23.21	$	&	$	32.5	$	\\
$	RFGC	286	$	&	$	16.28	$	&	$	70.31	$	&	$	67.30$	&	$		1.5	$	&	$	23.09	$	&	$	31.8	$	\\
$	RFGC	292	$	&	$	16.93	$	&	$	94.16	$	&	$	65.73$	&	$		1.19	$	&	$	23.4	$	&	$	28.4	$	\\
$	RFGC	296	$	&	$	16.53	$	&	$	90.36	$	&	$	69.10$	&	$		1.1	$	&	$	23.22	$	&	$	30.6	$	\\
$	RFGC	344	$	&	$	14.07	$	&	$	73.72	$	&	$	69.81$	&	$		1.4	$	&	$	23.39	$	&	$	30.4	$	\\
$	RFGC	357	$	&	$	15.46	$	&	$	82.21	$	&	$	63.82$	&	$		1.19	$	&	$	23.25	$	&	$	28.4	$	\\
$	RFGC	364	$	&	$	15.16	$	&	$	70.51	$	&	$	65.34$	&	$		1.5	$	&	$	22.92	$	&	$	31.4	$	\\
$	RFGC	374	$	&	$	14.32	$	&	$	88.24	$	&	$	70.60$	&	$		1.12	$	&	$	23.33	$	&	$	26.2	$	\\
$	RFGC	504	$	&	$	16.85	$	&	$	87.36	$	&	$	64.85$	&	$		1.32	$	&	$	22.98	$	&	$	24.4	$	\\
$	RFGC	505	$	&	$	14.09	$	&	$	66.08	$	&	$	68.44$	&	$		1.12	$	&	$	22.92	$	&	$	33.0	$	\\
$	RFGC	509	$	&	$	16.45	$	&	$	80.77	$	&	$	64.29$	&	$		1.47	$	&	$	22.95	$	&	$	33.5	$	\\
$	RFGC	510	$	&	$	16	$	&	$	77.24	$	&	$	67.64$	&	$		1.01	$	&	$	23.35	$	&	$	26.4	$	\\
$	RFGC	517	$	&	$	14.1	$	&	$	86.19	$	&	$	62.52$	&	$		1.32	$	&	$	23.33	$	&	$	30.1	$	\\
$	RFGC	531	$	&	$	16.96	$	&	$	69.28	$	&	$	71.77$	&	$		1.42	$	&	$	23.19	$	&	$	32.0	$	\\
$	RFGC	544	$	&	$	14.16	$	&	$	69.14	$	&	$	62.06$	&	$		1.33	$	&	$	23.38	$	&	$	25.3	$	\\
$	RFGC	560	$	&	$	16.58	$	&	$	87.55	$	&	$	68.57$	&	$		1.21	$	&	$	23.24	$	&	$	30.8	$	\\
$	RFGC	570	$	&	$	16.73	$	&	$	70.94	$	&	$	62.29$	&	$		1.06	$	&	$	23.01	$	&	$	31.9	$	\\
$	RFGC	582	$	&	$	14.7	$	&	$	75.21	$	&	$	62.13$	&	$		1.15	$	&	$	23.2	$	&	$	30.8	$	\\
$	RFGC	589	$	&	$	15.37	$	&	$	71.66	$	&	$	70.81$	&	$		1.34	$	&	$	23.12	$	&	$	31.5	$	\\
$	RFGC	598	$	&	$	15.59	$	&	$	89.32	$	&	$	64.13$	&	$		1.04	$	&	$	23.04	$	&	$	34.6	$	\\
$	RFGC	603	$	&	$	16.27	$	&	$	87.53	$	&	$	70.02$	&	$		1.34	$	&	$	23.04	$	&	$	32.4	$	\\
$	RFGC	605	$	&	$	14.29	$	&	$	80.19	$	&	$	68.62$	&	$		1.22	$	&	$	23.37	$	&	$	32.7	$	\\
$	RFGC	620	$	&	$	16.44	$	&	$	76.27	$	&	$	65.76$	&	$		1.25	$	&	$	23.06	$	&	$	26.8	$	\\
$	RFGC	622	$	&	$	14.07	$	&	$	91.75	$	&	$	70.14$	&	$		1.12	$	&	$	22.95	$	&	$	32.6	$	\\
$	RFGC	625	$	&	$	15.07	$	&	$	83.67	$	&	$	67.19$	&	$		1.3	$	&	$	23.37	$	&	$	26.8	$	\\
$	RFGC	627	$	&	$	14.98	$	&	$	87.58	$	&	$	68.57$	&	$		1.3	$	&	$	23.04	$	&	$	23.5	$	\\
$	RFGC	634	$	&	$	14.51	$	&	$	93.55	$	&	$	63.76$	&	$		1.04	$	&	$	23.25	$	&	$	34.3	$	\\
$	RFGC	642	$	&	$	14.49	$	&	$	71.23	$	&	$	68.50$	&	$		1.01	$	&	$	23.09	$	&	$	34.5	$	\\
$	RFGC	653	$	&	$	16.9	$	&	$	80.94	$	&	$	68.30$	&	$		1.34	$	&	$	23.37	$	&	$	23.5	$	\\
$	RFGC	674	$	&	$	14.27	$	&	$	73.71	$	&	$	63.63$	&	$		1.24	$	&	$	22.91	$	&	$	23.9	$	\\
$	RFGC	693	$	&	$	14.33	$	&	$	79.24	$	&	$	64.34$	&	$		1.09	$	&	$	23.32	$	&	$	32.7	$	\\
\hline
\end{tabular}
\end{table}

\begin{table}
\renewcommand\thetable{}
    \centering
	\caption{\textbf{A3} General Properties: Bulgeless spirals} 
    \label{Appendix : Table 3} 
    \begin{tabular}{lccccccc} 
    \hline
Name \footnote{Name of the galaxy from \cite{Karachentseva_2020}} sample & $M_B$ \footnote{B-band apparent magnitude} & Distance\footnote{Distance to galaxy in Mpc} & Vrot\footnote{asymptotic rotational velocity in km/s} & a/b \footnote{log(a/b) major to minor axes ration in 0.1 dec unit}& $\mu_B$ \footnote{Surface brightness in B band in $mag arcsec^{-1}$}&Diameter \footnote{Diameter of the galaxy in kpc} \\
\hline
$	NGC7816	$ & $	13.55	$ & $	76.56	$ & $	197.1	$ & $	0.03	$ & $	23.67	$ & $	36.96	$ \\
$	UGC00044	$ & $	16.78	$ & $	91.62	$ & $	100.4	$ & $	0.03	$ & $	25.25	$ & $	22.17	$ \\
$	UGC00048	$ & $	15.33	$ & $	65.46	$ & $	81.6	$ & $	0.04	$ & $	25.74	$ & $	31.6	$ \\
$	NGC7834	$ & $	14.64	$ & $	75.86	$ & $	115.3	$ & $	0.05	$ & $	23.84	$ & $	20.59	$ \\
$	NGC0039	$ & $	13.92	$ & $	72.44	$ & $	178.9	$ & $	0.05	$ & $	23.19	$ & $	23.11	$ \\
$	UGC00160	$ & $	16.26	$ & $	70.15	$ & $	265.1	$ & $	0.01	$ & $	25.76	$ & $	26.29	$ \\
$	IC1562	$ & $	13.46	$ & $	51.52	$ & $	62.5	$ & $	0.03	$ & $	23.18	$ & $	22.68	$ \\
$	NGC0198	$ & $	12.93	$ & $	76.21	$ & $	241.7	$ & $	0.04	$ & $	20.3	$ & $	28.56	$ \\
$	IC0043	$ & $	13.59	$ & $	71.78	$ & $	190.5	$ & $	0.04	$ & $	23.1	$ & $	26.29	$ \\
$	NGC0236	$ & $	14.24	$ & $	81.66	$ & $	115.8	$ & $	0.05	$ & $	23.39	$ & $	27.27	$ \\
$	NGC0255	$ & $	12.21	$ & $	21.88	$ & $	159.5	$ & $	0.04	$ & $	20.61	$ & $	12.99	$ \\
$	IC0056	$ & $	14.94	$ & $	86.7	$ & $	138	$ & $	0.05	$ & $	23.36	$ & $	20.5	$ \\
$	IC1666	$ & $	14.03	$ & $	72.44	$ & $	187.4	$ & $	0.03	$ & $	20.87	$ & $	19.67	$ \\
$	PGC005023	$ & $	15.63	$ & $	31.62	$ & $		$ & $	0	$ & $	24.05	$ & $	7.31	$ \\
$	UGC00929	$ & $	14.56	$ & $	107.65	$ & $	75.7	$ & $	0.04	$ & $	23.47	$ & $	31.31	$ \\
$	NGC0575	$ & $	13.39	$ & $	46.34	$ & $	166.5	$ & $	0.03	$ & $	23.2	$ & $	19.49	$ \\
$	NGC0628	$ & $	9.35	$ & $	10.14	$ & $	64.3	$ & $	0.03	$ & $	23.37	$ & $	29.49	$ \\
$	UGC01148	$ & $	14.71	$ & $	72.44	$ & $		$ & $	0	$ & $	24.09	$ & $	20.12	$ \\
$	UGC01347	$ & $	13.13	$ & $	82.04	$ & $	173	$ & $	0.02	$ & $	20.11	$ & $	22.79	$ \\
$	PGC007210	$ & $	14.78	$ & $	115.88	$ & $	286.7	$ & $	0.03	$ & $	23.48	$ & $	30.74	$ \\
$	UGC01478	$ & $	14	$ & $	70.79	$ & $	92.2	$ & $	0.05	$ & $	20.87	$ & $	17.53	$ \\
$	UGC01546	$ & $	14.34	$ & $	34.51	$ & $	87.9	$ & $	0.04	$ & $	23.11	$ & $	8.74	$ \\
$	PGC007942	$ & $	14.8	$ & $	74.82	$ & $	122.9	$ & $	0.04	$ & $	23.97	$ & $	24.99	$ \\
$	UGC02094	$ & $	13.41	$ & $	75.51	$ & $	285.1	$ & $	0.04	$ & $	20.75	$ & $	25.81	$ \\
$	UGC02174	$ & $	14.47	$ & $	75.51	$ & $	160.6	$ & $	0.01	$ & $	25.3	$ & $	43.83	$ \\
$	ESO546-011	$ & $	14.28	$ & $	63.68	$ & $	72.6	$ & $	0.03	$ & $	23.43	$ & $	20.78	$ \\
$	NGC1067	$ & $	13.68	$ & $	66.37	$ & $	189.6	$ & $	0.02	$ & $	23.17	$ & $	18.87	$ \\
$	UGC02323	$ & $	14.5	$ & $	116.95	$ & $	164.7	$ & $	0.03	$ & $	24.04	$ & $	29.63	$ \\
$	UGC02623	$ & $	14.5	$ & $	65.16	$ & $	120.8	$ & $	0.03	$ & $	25.85	$ & $	39.6	$ \\
$	UGC02692	$ & $	13.66	$ & $	89.54	$ & $	208.1	$ & $	0.04	$ & $	20.46	$ & $	22.68	$ \\
$	UGC02712	$ & $	15.72	$ & $	101.86	$ & $		$ & $	0	$ & $	26.46	$ & $	40.9	$ \\
$	UGC02721	$ & $	16.25	$ & $	92.47	$ & $	57.6	$ & $	0.04	$ & $	25.15	$ & $	21.86	$ \\
$	NGC1325A	$ & $	13.23	$ & $	16.6	$ & $	52.5	$ & $	0.03	$ & $	23.48	$ & $	9.2	$ \\
$	ESO548-035	$ & $	13.15	$ & $	58.08	$ & $	141.7	$ & $	0.04	$ & $	20.86	$ & $	23.32	$ \\
$	NGC1376	$ & $	12.61	$ & $	58.08	$ & $	273	$ & $	0.01	$ & $	20.91	$ & $	31.46	$ \\
$	IC0342	$ & $	6.14	$ & $	3.44	$ & $	230.4	$ & $	0.02	$ & $	24.85	$ & $	19.94	$ \\
$	UGC02859	$ & $	14.02	$ & $	80.54	$ & $	243.8	$ & $	0.03	$ & $	24.23	$ & $	19.04	$ \\
$	UGC03051	$ & $	14.98	$ & $	99.08	$ & $	244.2	$ & $	0.03	$ & $	24.46	$ & $	25.69	$ \\
$	NGC1599	$ & $	13.87	$ & $	55.72	$ & $	127.6	$ & $	0.02	$ & $	20.35	$ & $	13.17	$ \\
$	IC0391	$ & $	12.39	$ & $	26.55	$ & $	179.1	$ & $	0.02	$ & $	20.41	$ & $	10.66	$ \\
$	PGC017323	$ & $	13.1	$ & $	28.84	$ & $	100.8	$ & $	0.03	$ & $	23.94	$ & $	18.78	$ \\
$	PGC018031	$ & $	13.69	$ & $	101.39	$ & $	206.2	$ & $	0.03	$ & $	23.4	$ & $	28.17	$ \\
$	IC0441	$ & $	13.16	$ & $	29.38	$ & $	125.3	$ & $	0.03	$ & $	23.5	$ & $	10.51	$ \\
$	UGC03574	$ & $	13.53	$ & $	18.54	$ & $	166.8	$ & $	0.03	$ & $	23.35	$ & $	7.97	$ \\
$	UGC03701	$ & $	14.83	$ & $	45.92	$ & $		$ & $	0	$ & $	25.06	$ & $	23.21	$ \\
$	UGC03806	$ & $	14.77	$ & $	79.43	$ & $	250.8	$ & $	0.03	$ & $	23.69	$ & $	20.59	$ \\
$	UGC03825	$ & $	14.61	$ & $	121.9	$ & $	229.9	$ & $	0.04	$ & $	24.07	$ & $	41.66	$ \\
$	UGC03886	$ & $	15.67	$ & $	73.45	$ & $	175.7	$ & $	0.02	$ & $	24.8	$ & $	22.37	$ \\
$	UGC03924	$ & $	15.47	$ & $	74.13	$ & $	116.1	$ & $	0.03	$ & $	24.15	$ & $	19.67	$ \\
$	NGC2500	$ & $	12.01	$ & $	9.77	$ & $	130.5	$ & $	0.02	$ & $	20.88	$ & $	6.98	$ \\
$	NGC2514	$ & $	13.8	$ & $	70.15	$ & $	145.6	$ & $	0.03	$ & $	23.09	$ & $	24.53	$ \\
$	PGC086610	$ & $	16.18	$ & $	70.15	$ & $		$ & $	0	$ & $	24.84	$ & $	17.77	$ \\
$	PGC023378	$ & $	14.22	$ & $	64.57	$ & $	140.8	$ & $	0.03	$ & $	23.37	$ & $	20.59	$ \\
$	UGC04380	$ & $	14.71	$ & $	111.69	$ & $	144.5	$ & $	0.03	$ & $	23.5	$ & $	29.63	$ \\
$	UGC04445	$ & $	14.69	$ & $	94.62	$ & $	66.5	$ & $	0.02	$ & $	23.76	$ & $	27.52	$ \\
$	IC0509	$ & $	13.55	$ & $	80.17	$ & $	80.3	$ & $	0.05	$ & $	20.5	$ & $	23.86	$ \\
$	NGC2607	$ & $	14.71	$ & $	51.76	$ & $	84	$ & $	0.02	$ & $	23.26	$ & $	12.52	$ \\
$	NGC2661	$ & $	13.64	$ & $	59.43	$ & $	98.3	$ & $	0.04	$ & $	23.37	$ & $	25.57	$ \\
$	UGC04853	$ & $	14.95	$ & $	38.73	$ & $	61.5	$ & $	0.05	$ & $	23.78	$ & $	10.76	$ \\
$	UGC05015	$ & $	15.51	$ & $	25.35	$ & $	148.4	$ & $	0.02	$ & $	25.35	$ & $	11.96	$ \\
$	NGC2967	$ & $	11.79	$ & $	27.04	$ & $	165.5	$ & $	0.03	$ & $	20.6	$ & $	16.82	$ \\
$	UGC05274	$ & $	14.69	$ & $	86.3	$ & $	261.5	$ & $	0.01	$ & $	23.53	$ & $	23.97	$ \\
$	ESO566-019	$ & $	13.74	$ & $	51.76	$ & $	253.7	$ & $	0.02	$ & $	23.78	$ & $	24.99	$ \\
$	ESO499-011	$ & $	14.7	$ & $	35.81	$ & $	114.2	$ & $	0.02	$ & $	23.64	$ & $	10.18	$ \\
$	PGC028556	$ & $	14.52	$ & $	106.66	$ & $	63.6	$ & $	0.04	$ & $	23.1	$ & $	25.81	$ \\
$	ESO567-010	$ & $	14.08	$ & $	42.27	$ & $	61.3	$ & $	0.02	$ & $	23.7	$ & $	15.84	$ \\
\hline
\end{tabular}
\end{table}

\begin{table}
\renewcommand\thetable{}
\centering
\begin{minipage}{230mm}
{\small
\hfill{}
\hspace*{-10.00 cm}
\caption{\textbf{A4} Input for MAGPHYS: LSBs }
\begin{tabular}{|c|c|c|c|c|c|c|c|c|c|c|}
\hline
id & FUV(Jy) & NUV(Jy) & SDSS u(Jy)& SDSS g(Jy) & SDSS r(Jy) & SDSS i(Jy) & SDSS z(Jy) & J(Jy) & H(Jy) & Ks(Jy) \\  
\hline
$	1048	$ & $	6.3E-07	$ & $	9.0E-07	$ & $	1.3E-06	$ & $	2.5E-06	$ & $	2.9E-06	$ & $	3.3E-06	$ & $	3.7E-06	$ & $	1.3E-06	$ & $	1.6E-06	$ & $	5.5E-08	$ \\
$	1084	$ & $	4.6E-08	$ & $	5.1E-08	$ & $	6.6E-08	$ & $	1.3E-07	$ & $	1.7E-07	$ & $	2.0E-07	$ & $	2.3E-07	$ & $	7.1E-08	$ & $	5.7E-08	$ & $	2.1E-07	$ \\
$	1141	$ & $	1.6E-06	$ & $	2.1E-06	$ & $	5.7E-06	$ & $	1.6E-05	$ & $	2.5E-05	$ & $	3.0E-05	$ & $	3.5E-05	$ & $	8.4E-06	$ & $	2.8E-05	$ & $	1.4E-05	$ \\
$	1010	$ & $	5.5E-07	$ & $	7.1E-07	$ & $	1.5E-06	$ & $	4.9E-06	$ & $	8.3E-06	$ & $	1.1E-05	$ & $	1.4E-05	$ & $	4.7E-06	$ & $	2.3E-06	$ & $	3.0E-08	$ \\
$	1025	$ & $	1.1E-07	$ & $	1.5E-07	$ & $	2.7E-07	$ & $	8.1E-07	$ & $	1.4E-06	$ & $	2.0E-06	$ & $	2.6E-06	$ & $	9.9E-07	$ & $	6.1E-07	$ & $	7.9E-07	$ \\
$	1085	$ & $	2.5E-07	$ & $	3.0E-07	$ & $	5.0E-07	$ & $	1.2E-06	$ & $	1.7E-06	$ & $	2.0E-06	$ & $	2.3E-06	$ & $	1.1E-06	$ & $	3.5E-07	$ & $	9.2E-07	$ \\
$	1093	$ & $	1.7E-07	$ & $	2.8E-07	$ & $	6.3E-07	$ & $	1.6E-06	$ & $	1.8E-06	$ & $	2.1E-06	$ & $	2.3E-06	$ & $	1.7E-06	$ & $	1.6E-06	$ & $	1.0E-06	$ \\
$	1013	$ & $	8.8E-07	$ & $	1.1E-06	$ & $	1.8E-06	$ & $	4.5E-06	$ & $	6.2E-06	$ & $	7.5E-06	$ & $	8.8E-06	$ & $	1.2E-06	$ & $	4.4E-06	$ & $	1.0E-06	$ \\
$	1039	$ & $	3.2E-06	$ & $	3.3E-06	$ & $	4.1E-06	$ & $	9.2E-06	$ & $	1.4E-05	$ & $	1.7E-05	$ & $	2.2E-05	$ & $	2.1E-06	$ & $	5.4E-06	$ & $	9.8E-06	$ \\
$	1181	$ & $	4.5E-06	$ & $	6.0E-06	$ & $	8.9E-06	$ & $	2.0E-05	$ & $	2.9E-05	$ & $	3.6E-05	$ & $	4.4E-05	$ & $	1.9E-06	$ & $	1.1E-05	$ & $	3.7E-05	$ \\
$	1094	$ & $	1.5E-08	$ & $	2.5E-08	$ & $	7.3E-08	$ & $	2.2E-07	$ & $	3.4E-07	$ & $	4.4E-07	$ & $	5.2E-07	$ & $	2.0E-07	$ & $	3.1E-07	$ & $	3.1E-07	$ \\
$	1142	$ & $	4.6E-07	$ & $	8.6E-07	$ & $	2.0E-06	$ & $	5.8E-06	$ & $	9.1E-06	$ & $	1.2E-05	$ & $	1.5E-05	$ & $	8.0E-06	$ & $	7.2E-06	$ & $	1.6E-06	$ \\
$	1071	$ & $	1.8E-07	$ & $	4.2E-07	$ & $	1.4E-06	$ & $	5.2E-06	$ & $	9.5E-06	$ & $	1.3E-05	$ & $	1.8E-05	$ & $	7.2E-06	$ & $	3.8E-07	$ & $	6.4E-06	$ \\
$	1133	$ & $	2.6E-08	$ & $	3.0E-08	$ & $	6.6E-08	$ & $	2.3E-07	$ & $	4.6E-07	$ & $	6.4E-07	$ & $	8.5E-07	$ & $	3.3E-07	$ & $	3.2E-07	$ & $	4.7E-07	$ \\
$	1057	$ & $	1.1E-07	$ & $	2.4E-07	$ & $	8.6E-07	$ & $	3.1E-06	$ & $	5.7E-06	$ & $	7.9E-06	$ & $	1.0E-05	$ & $	2.6E-07	$ & $	6.1E-06	$ & $	5.4E-06	$ \\
$	1097	$ & $	4.2E-07	$ & $	7.1E-07	$ & $	2.0E-06	$ & $	7.2E-06	$ & $	1.4E-05	$ & $	2.0E-05	$ & $	2.7E-05	$ & $	1.7E-06	$ & $	3.9E-06	$ & $	9.8E-06	$ \\
$	1116	$ & $	6.4E-07	$ & $	7.8E-07	$ & $	1.1E-06	$ & $	2.5E-06	$ & $	3.6E-06	$ & $	4.5E-06	$ & $	5.6E-06	$ & $	3.0E-06	$ & $	3.8E-07	$ & $	3.2E-06	$ \\
$	1083	$ & $	1.8E-07	$ & $	2.7E-07	$ & $	4.5E-07	$ & $	9.5E-07	$ & $	1.3E-06	$ & $	1.6E-06	$ & $	1.9E-06	$ & $	1.3E-06	$ & $	2.4E-07	$ & $	1.1E-06	$ \\
$	1179	$ & $	1.6E-07	$ & $	2.9E-07	$ & $	6.4E-07	$ & $	1.7E-06	$ & $	2.7E-06	$ & $	3.6E-06	$ & $	4.5E-06	$ & $	2.9E-07	$ & $	3.4E-06	$ & $	4.0E-06	$ \\
$	1155	$ & $	1.4E-07	$ & $	3.2E-07	$ & $	9.7E-07	$ & $	3.1E-06	$ & $	5.4E-06	$ & $	7.4E-06	$ & $	9.9E-06	$ & $	3.7E-06	$ & $	1.2E-07	$ & $	2.0E-06	$ \\
$	1098	$ & $	1.0E-06	$ & $	1.5E-06	$ & $	3.7E-06	$ & $	1.2E-05	$ & $	2.2E-05	$ & $	3.0E-05	$ & $	4.0E-05	$ & $	4.7E-06	$ & $	2.4E-05	$ & $	1.4E-05	$ \\
$	1110	$ & $	9.3E-08	$ & $	1.2E-07	$ & $	2.3E-07	$ & $	6.6E-07	$ & $	1.1E-06	$ & $	1.5E-06	$ & $	2.0E-06	$ & $	8.3E-07	$ & $	7.9E-07	$ & $	8.3E-07	$ \\
$	1178	$ & $	2.4E-06	$ & $	2.9E-06	$ & $	4.5E-06	$ & $	1.2E-05	$ & $	1.9E-05	$ & $	2.4E-05	$ & $	3.1E-05	$ & $	8.1E-06	$ & $	1.4E-05	$ & $	2.1E-05	$ \\
$	1143	$ & $	4.6E-07	$ & $	5.4E-07	$ & $	7.5E-07	$ & $	1.6E-06	$ & $	2.2E-06	$ & $	2.7E-06	$ & $	3.2E-06	$ & $	1.5E-06	$ & $	1.8E-06	$ & $	6.0E-07	$ \\
$	1024	$ & $	1.5E-06	$ & $	2.5E-06	$ & $	4.2E-06	$ & $	1.0E-05	$ & $	1.5E-05	$ & $	2.0E-05	$ & $	2.5E-05	$ & $	4.1E-06	$ & $	1.5E-05	$ & $	1.7E-05	$ \\
$	1159	$ & $	3.0E-07	$ & $	6.5E-07	$ & $	1.9E-06	$ & $	6.6E-06	$ & $	1.2E-05	$ & $	1.6E-05	$ & $	2.2E-05	$ & $	9.1E-06	$ & $	1.2E-05	$ & $	1.7E-05	$ \\
$	1101	$ & $	5.5E-08	$ & $	7.2E-08	$ & $	1.2E-07	$ & $	2.7E-07	$ & $	3.9E-07	$ & $	4.9E-07	$ & $	5.9E-07	$ & $	3.2E-07	$ & $	9.1E-08	$ & $	7.4E-08	$ \\
$	1050	$ & $	1.7E-06	$ & $	2.8E-06	$ & $	7.4E-06	$ & $	2.3E-05	$ & $	3.9E-05	$ & $	5.0E-05	$ & $	6.3E-05	$ & $	1.4E-05	$ & $	1.7E-05	$ & $	3.7E-05	$ \\
$	1162	$ & $	4.4E-07	$ & $	7.0E-07	$ & $	2.2E-06	$ & $	7.2E-06	$ & $	1.3E-05	$ & $	1.7E-05	$ & $	2.0E-05	$ & $	8.4E-06	$ & $	4.3E-06	$ & $	1.2E-05	$ \\
$	1038	$ & $	4.8E-07	$ & $	6.5E-07	$ & $	8.4E-07	$ & $	2.0E-06	$ & $	3.8E-06	$ & $	5.5E-06	$ & $	7.4E-06	$ & $	2.5E-06	$ & $	3.3E-06	$ & $	5.8E-06	$ \\
$	1148	$ & $	7.9E-07	$ & $	1.2E-06	$ & $	2.4E-06	$ & $	6.7E-06	$ & $	1.2E-05	$ & $	1.6E-05	$ & $	2.1E-05	$ & $	2.2E-06	$ & $	1.3E-05	$ & $	1.9E-07	$ \\
$	1032	$ & $	7.0E-07	$ & $	8.5E-07	$ & $	1.4E-06	$ & $	3.9E-06	$ & $	7.0E-06	$ & $	9.6E-06	$ & $	1.3E-05	$ & $	2.2E-06	$ & $	6.1E-06	$ & $	3.7E-07	$ \\
$	1125	$ & $	9.0E-07	$ & $	1.4E-06	$ & $	3.8E-06	$ & $	1.2E-05	$ & $	2.0E-05	$ & $	2.5E-05	$ & $	3.2E-05	$ & $	1.2E-06	$ & $	2.5E-05	$ & $	1.5E-05	$ \\
$	1154	$ & $	1.9E-07	$ & $	3.1E-07	$ & $	1.2E-06	$ & $	4.6E-06	$ & $	8.9E-06	$ & $	1.2E-05	$ & $	1.6E-05	$ & $	4.5E-06	$ & $	3.7E-07	$ & $	1.0E-05	$ \\
$	1158	$ & $	1.5E-06	$ & $	2.4E-06	$ & $	5.3E-06	$ & $	1.6E-05	$ & $	2.9E-05	$ & $	4.0E-05	$ & $	5.4E-05	$ & $	8.5E-06	$ & $	3.9E-06	$ & $	2.0E-05	$ \\
$	1117	$ & $	2.6E-06	$ & $	3.0E-06	$ & $	4.0E-06	$ & $	8.7E-06	$ & $	1.1E-05	$ & $	1.3E-05	$ & $	1.5E-05	$ & $	7.0E-06	$ & $	1.1E-05	$ & $	5.4E-06	$ \\
$	1008	$ & $	1.4E-07	$ & $	2.2E-07	$ & $	8.3E-07	$ & $	3.1E-06	$ & $	5.9E-06	$ & $	8.2E-06	$ & $	1.1E-05	$ & $	1.0E-06	$ & $	1.4E-06	$ & $	3.0E-06	$ \\
$	1014	$ & $	3.9E-06	$ & $	4.1E-06	$ & $	5.1E-06	$ & $	1.1E-05	$ & $	1.7E-05	$ & $	2.1E-05	$ & $	2.7E-05	$ & $	1.6E-05	$ & $	7.3E-06	$ & $	4.6E-06	$ \\
$	1090	$ & $	1.2E-06	$ & $	1.5E-06	$ & $	2.3E-06	$ & $	5.3E-06	$ & $	7.3E-06	$ & $	8.9E-06	$ & $	1.1E-05	$ & $	5.5E-06	$ & $	4.0E-06	$ & $	7.5E-06	$ \\
$	1066	$ & $	1.9E-07	$ & $	3.6E-07	$ & $	1.3E-06	$ & $	4.1E-06	$ & $	6.8E-06	$ & $	8.8E-06	$ & $	1.0E-05	$ & $	4.8E-07	$ & $	6.9E-06	$ & $	6.4E-08	$ \\
$	1106	$ & $	2.1E-07	$ & $	3.4E-07	$ & $	8.2E-07	$ & $	2.3E-06	$ & $	3.3E-06	$ & $	4.1E-06	$ & $	4.9E-06	$ & $	2.8E-06	$ & $	1.5E-06	$ & $	8.1E-07	$ \\
$	1076	$ & $	7.5E-07	$ & $	1.0E-06	$ & $	2.9E-06	$ & $	8.5E-06	$ & $	1.4E-05	$ & $	1.7E-05	$ & $	2.0E-05	$ & $	8.7E-06	$ & $	1.5E-05	$ & $	7.0E-06	$ \\
$	1170	$ & $	3.9E-06	$ & $	4.9E-06	$ & $	6.8E-06	$ & $	1.5E-05	$ & $	2.1E-05	$ & $	2.6E-05	$ & $	3.2E-05	$ & $	1.0E-05	$ & $	1.5E-05	$ & $	1.1E-06	$ \\
$	1001	$ & $	4.4E-07	$ & $	6.8E-07	$ & $	2.7E-06	$ & $	9.6E-06	$ & $	1.8E-05	$ & $	2.4E-05	$ & $	3.1E-05	$ & $	8.1E-06	$ & $	2.1E-05	$ & $	2.3E-05	$ \\
$	1118	$ & $	3.1E-07	$ & $	5.1E-07	$ & $	8.1E-07	$ & $	1.9E-06	$ & $	2.6E-06	$ & $	3.2E-06	$ & $	4.0E-06	$ & $	2.3E-07	$ & $	2.9E-06	$ & $	3.6E-06	$ \\
$	1064	$ & $	1.6E-06	$ & $	2.6E-06	$ & $	5.3E-06	$ & $	1.5E-05	$ & $	2.7E-05	$ & $	3.6E-05	$ & $	4.7E-05	$ & $	1.8E-05	$ & $	2.7E-05	$ & $	5.1E-06	$ \\
$	1124	$ & $	1.9E-08	$ & $	2.8E-08	$ & $	5.7E-08	$ & $	1.4E-07	$ & $	1.9E-07	$ & $	2.3E-07	$ & $	2.6E-07	$ & $	1.3E-07	$ & $	1.6E-07	$ & $	2.6E-07	$ \\
$	1134	$ & $	2.2E-07	$ & $	2.6E-07	$ & $	4.0E-07	$ & $	1.1E-06	$ & $	1.9E-06	$ & $	2.6E-06	$ & $	3.4E-06	$ & $	8.8E-08	$ & $	1.0E-06	$ & $	5.4E-07	$ \\
$	1089	$ & $	1.7E-06	$ & $	2.0E-06	$ & $	2.6E-06	$ & $	5.4E-06	$ & $	7.3E-06	$ & $	8.8E-06	$ & $	1.0E-05	$ & $	5.6E-06	$ & $	2.2E-06	$ & $	9.2E-06	$ \\
$	1163	$ & $	2.7E-07	$ & $	5.0E-07	$ & $	1.3E-06	$ & $	3.7E-06	$ & $	5.9E-06	$ & $	7.6E-06	$ & $	9.5E-06	$ & $	3.7E-07	$ & $	1.7E-06	$ & $	3.3E-06	$ \\
$	1166	$ & $	2.4E-07	$ & $	3.9E-07	$ & $	8.3E-07	$ & $	2.3E-06	$ & $	3.7E-06	$ & $	4.8E-06	$ & $	6.0E-06	$ & $	1.2E-06	$ & $	1.9E-06	$ & $	2.0E-06	$ \\
$	1088	$ & $	1.1E-06	$ & $	1.2E-06	$ & $	1.6E-06	$ & $	3.8E-06	$ & $	5.6E-06	$ & $	7.1E-06	$ & $	8.7E-06	$ & $	8.9E-07	$ & $	2.4E-06	$ & $	7.5E-06	$ \\
$	1152	$ & $	2.8E-07	$ & $	1.1E-06	$ & $	1.1E-05	$ & $	3.8E-05	$ & $	7.2E-05	$ & $	9.4E-05	$ & $	1.1E-04	$ & $	3.7E-05	$ & $	8.4E-05	$ & $	9.2E-05	$ \\
$	1063	$ & $	1.9E-06	$ & $	2.9E-06	$ & $	5.3E-06	$ & $	1.5E-05	$ & $	2.5E-05	$ & $	3.3E-05	$ & $	4.3E-05	$ & $	6.9E-06	$ & $	1.0E-05	$ & $	1.9E-05	$ \\
$	1069	$ & $	1.1E-06	$ & $	1.8E-06	$ & $	3.2E-06	$ & $	8.3E-06	$ & $	1.3E-05	$ & $	1.7E-05	$ & $	2.2E-05	$ & $	1.1E-05	$ & $	4.0E-06	$ & $	1.7E-05	$ \\
$	1102	$ & $	1.1E-06	$ & $	1.4E-06	$ & $	3.0E-06	$ & $	1.0E-05	$ & $	2.0E-05	$ & $	2.8E-05	$ & $	3.8E-05	$ & $	1.2E-05	$ & $	2.0E-05	$ & $	3.9E-06	$ \\
$	1147	$ & $	2.3E-06	$ & $	3.5E-06	$ & $	6.7E-06	$ & $	1.8E-05	$ & $	2.8E-05	$ & $	3.5E-05	$ & $	4.3E-05	$ & $	5.7E-06	$ & $	1.5E-05	$ & $	2.5E-05	$ \\
$	1135	$ & $	3.5E-07	$ & $	5.0E-07	$ & $	1.1E-06	$ & $	3.0E-06	$ & $	4.7E-06	$ & $	5.9E-06	$ & $	7.2E-06	$ & $	9.4E-07	$ & $	4.5E-06	$ & $	3.1E-06	$ \\
$	1130	$ & $	1.1E-07	$ & $	1.7E-07	$ & $	3.4E-07	$ & $	8.6E-07	$ & $	1.1E-06	$ & $	1.3E-06	$ & $	1.5E-06	$ & $	7.3E-07	$ & $	7.2E-07	$ & $	5.7E-07	$ \\
$	1044	$ & $	1.1E-06	$ & $	1.5E-06	$ & $	2.9E-06	$ & $	7.6E-06	$ & $	1.1E-05	$ & $	1.3E-05	$ & $	1.6E-05	$ & $	6.5E-06	$ & $	7.5E-06	$ & $	2.8E-06	$ \\
$	1119	$ & $	4.1E-06	$ & $	5.0E-06	$ & $	7.9E-06	$ & $	2.3E-05	$ & $	4.1E-05	$ & $	5.7E-05	$ & $	7.7E-05	$ & $	1.2E-05	$ & $	3.2E-05	$ & $	1.2E-05	$ \\
$	1016	$ & $	6.6E-07	$ & $	9.7E-07	$ & $	1.6E-06	$ & $	3.8E-06	$ & $	5.6E-06	$ & $	7.0E-06	$ & $	8.6E-06	$ & $	9.8E-07	$ & $	6.5E-06	$ & $	5.2E-07	$ \\
$	1120	$ & $	3.2E-07	$ & $	4.0E-07	$ & $	4.8E-07	$ & $	7.8E-07	$ & $	9.2E-07	$ & $	1.1E-06	$ & $	1.2E-06	$ & $	3.6E-07	$ & $	2.8E-07	$ & $	1.1E-06	$ \\
$	1046	$ & $	1.1E-08	$ & $	4.6E-08	$ & $	6.4E-07	$ & $	2.6E-06	$ & $	5.1E-06	$ & $	7.0E-06	$ & $	8.9E-06	$ & $	1.6E-07	$ & $	4.2E-06	$ & $	4.8E-06	$ \\
\hline
\end{tabular}}
\hfill{}
\end{minipage}
\end{table}

\begin{table}
\renewcommand\thetable{}
\centering
\begin{minipage}{230mm}
{\small
\hfill{}
\hspace*{-10.00 cm}
\caption{\textbf{A5} Input for MAGPHYS: STs }
\begin{tabular}{|c|c|c|c|c|c|c|c|c|c|c|}
\hline
id & FUV \footnote{GALEX FUV flux density in units of Jansky}& NUV\footnote{GALEX flux density (Jansky)} & u\footnote{SDSS u band flux density (Jansky)}& g \footnote{SDSS g band flux density (Jansky)}& r\footnote{SDSS r band flux density (Jansky)} & i \footnote{i flux density (Jansky)} & z \footnote{z flux density (Jansky)} & J \footnote{2MASS J band flux density (Jansky)} & H \footnote{2MASS H flux density (Jansky)} & Ks \footnote{2MASS Ks flux density (Jansky)} \\  
\hline
$	1115	$ & $	3.5E-07	$ & $	5.7E-07	$ & $	1.5E-06	$ & $	5.3E-06	$ & $	9.9E-06	$ & $	1.4E-05	$ & $	1.8E-05	$ & $	5.4E-06	$ & $	1.2E-05	$ & $	2.3E-07	$ \\
$	1076	$ & $	3.6E-07	$ & $	5.4E-07	$ & $	2.0E-06	$ & $	6.4E-06	$ & $	1.1E-05	$ & $	1.4E-05	$ & $	1.7E-05	$ & $	1.8E-07	$ & $	1.1E-05	$ & $	4.1E-08	$ \\
$	1001	$ & $	2.9E-07	$ & $	4.4E-07	$ & $	1.1E-06	$ & $	3.5E-06	$ & $	6.2E-06	$ & $	8.4E-06	$ & $	1.1E-05	$ & $	1.1E-06	$ & $	3.0E-06	$ & $	8.8E-07	$ \\
$	1118	$ & $	1.7E-07	$ & $	3.5E-07	$ & $	1.6E-06	$ & $	5.4E-06	$ & $	9.7E-06	$ & $	1.3E-05	$ & $	1.5E-05	$ & $	7.7E-07	$ & $	5.2E-06	$ & $	3.9E-07	$ \\
$	1064	$ & $	1.4E-06	$ & $	2.3E-06	$ & $	7.2E-06	$ & $	2.1E-05	$ & $	3.4E-05	$ & $	4.1E-05	$ & $	4.7E-05	$ & $	2.4E-05	$ & $	2.7E-05	$ & $	7.3E-06	$ \\
$	1047	$ & $	8.8E-08	$ & $	2.0E-07	$ & $	8.4E-07	$ & $	3.1E-06	$ & $	5.7E-06	$ & $	7.9E-06	$ & $	1.0E-05	$ & $	2.2E-06	$ & $	1.8E-07	$ & $	2.1E-06	$ \\
$	1030	$ & $	2.0E-07	$ & $	3.5E-07	$ & $	2.6E-06	$ & $	1.1E-05	$ & $	2.3E-05	$ & $	3.3E-05	$ & $	4.3E-05	$ & $	1.1E-05	$ & $	1.5E-05	$ & $	5.3E-06	$ \\
$	1089	$ & $	3.7E-07	$ & $	6.3E-07	$ & $	2.0E-06	$ & $	6.2E-06	$ & $	1.0E-05	$ & $	1.3E-05	$ & $	1.5E-05	$ & $	1.2E-06	$ & $	2.1E-06	$ & $	2.0E-07	$ \\
$	1006	$ & $	3.3E-08	$ & $	2.8E-07	$ & $	1.7E-06	$ & $	6.9E-06	$ & $	1.1E-05	$ & $	1.4E-05	$ & $	1.8E-05	$ & $	1.5E-06	$ & $	3.0E-06	$ & $	1.2E-06	$ \\
$	1063	$ & $	2.5E-07	$ & $	4.6E-07	$ & $	1.7E-06	$ & $	5.5E-06	$ & $	9.7E-06	$ & $	1.3E-05	$ & $	1.6E-05	$ & $	5.5E-06	$ & $	4.0E-06	$ & $	4.4E-06	$ \\
$	1069	$ & $	5.7E-07	$ & $	9.7E-07	$ & $	2.3E-06	$ & $	7.6E-06	$ & $	1.4E-05	$ & $	1.9E-05	$ & $	2.6E-05	$ & $	1.1E-05	$ & $	1.1E-05	$ & $	8.0E-06	$ \\
$	1049	$ & $	7.8E-07	$ & $	1.5E-06	$ & $	2.8E-06	$ & $	6.7E-06	$ & $	9.2E-06	$ & $	1.1E-05	$ & $	1.4E-05	$ & $	6.2E-07	$ & $	2.5E-06	$ & $	1.6E-07	$ \\
$	1102	$ & $	7.2E-07	$ & $	1.3E-06	$ & $	4.2E-06	$ & $	1.4E-05	$ & $	2.1E-05	$ & $	2.6E-05	$ & $	3.2E-05	$ & $	6.7E-06	$ & $	4.6E-06	$ & $	4.9E-06	$ \\
$	1103	$ & $	4.0E-07	$ & $	6.2E-07	$ & $	2.2E-06	$ & $	7.3E-06	$ & $	1.3E-05	$ & $	1.7E-05	$ & $	2.0E-05	$ & $	7.8E-06	$ & $	5.7E-06	$ & $	3.3E-06	$ \\
$	1044	$ & $	2.6E-07	$ & $	4.4E-07	$ & $	1.1E-06	$ & $	3.7E-06	$ & $	7.5E-06	$ & $	1.1E-05	$ & $	1.5E-05	$ & $	2.3E-06	$ & $	7.5E-06	$ & $	1.9E-07	$ \\
$	1119	$ & $	1.5E-08	$ & $	3.5E-08	$ & $	1.8E-07	$ & $	5.5E-07	$ & $	9.1E-07	$ & $	1.1E-06	$ & $	1.3E-06	$ & $	2.9E-07	$ & $	2.8E-07	$ & $	2.7E-07	$ \\
$	1029	$ & $	2.1E-07	$ & $	3.2E-07	$ & $	8.4E-07	$ & $	3.0E-06	$ & $	5.7E-06	$ & $	8.1E-06	$ & $	1.1E-05	$ & $	2.4E-06	$ & $	3.9E-06	$ & $	1.1E-06	$ \\
$	1046	$ & $	1.8E-07	$ & $	3.7E-07	$ & $	1.3E-06	$ & $	5.1E-06	$ & $	1.0E-05	$ & $	1.5E-05	$ & $	2.0E-05	$ & $	7.7E-06	$ & $	1.2E-05	$ & $	3.7E-06	$ \\
$	1018	$ & $	3.0E-07	$ & $	4.7E-07	$ & $	9.4E-07	$ & $	2.5E-06	$ & $	3.2E-06	$ & $	3.7E-06	$ & $	4.3E-06	$ & $	2.1E-06	$ & $	1.6E-06	$ & $	1.7E-07	$ \\
$	1042	$ & $	3.4E-07	$ & $	4.4E-07	$ & $	7.7E-07	$ & $	2.3E-06	$ & $	4.2E-06	$ & $	5.7E-06	$ & $	7.5E-06	$ & $	2.6E-06	$ & $	1.5E-06	$ & $	2.4E-06	$ \\
$	1113	$ & $	6.1E-08	$ & $	1.6E-07	$ & $	4.8E-07	$ & $	1.5E-06	$ & $	2.6E-06	$ & $	3.8E-06	$ & $	5.2E-06	$ & $	1.6E-06	$ & $	1.6E-07	$ & $	1.5E-06	$ \\
$	1053	$ & $	3.6E-07	$ & $	5.3E-07	$ & $	1.3E-06	$ & $	3.7E-06	$ & $	5.9E-06	$ & $	7.4E-06	$ & $	9.1E-06	$ & $	1.5E-06	$ & $	1.4E-06	$ & $	5.3E-07	$ \\
$	1015	$ & $	2.3E-07	$ & $	3.7E-07	$ & $	8.4E-07	$ & $	2.7E-06	$ & $	4.4E-06	$ & $	5.8E-06	$ & $	7.4E-06	$ & $	8.0E-07	$ & $	2.8E-06	$ & $	1.6E-07	$ \\
$	1075	$ & $	1.9E-08	$ & $	4.2E-08	$ & $	7.9E-07	$ & $	3.7E-06	$ & $	7.8E-06	$ & $	1.1E-05	$ & $	1.5E-05	$ & $	6.9E-06	$ & $	6.9E-06	$ & $	5.7E-06	$ \\
$	1031	$ & $	5.4E-07	$ & $	9.4E-07	$ & $	3.2E-06	$ & $	1.1E-05	$ & $	1.9E-05	$ & $	2.5E-05	$ & $	3.1E-05	$ & $	9.4E-06	$ & $	6.5E-06	$ & $	4.7E-06	$ \\
$	1036	$ & $	3.3E-07	$ & $	5.2E-07	$ & $	8.7E-07	$ & $	2.0E-06	$ & $	2.4E-06	$ & $	2.7E-06	$ & $	3.0E-06	$ & $	2.7E-07	$ & $	2.2E-06	$ & $	1.1E-07	$ \\
$	1104	$ & $	1.8E-07	$ & $	2.5E-07	$ & $	4.3E-07	$ & $	1.0E-06	$ & $	1.4E-06	$ & $	1.6E-06	$ & $	1.9E-06	$ & $	9.0E-08	$ & $	9.1E-07	$ & $	7.8E-08	$ \\
$	1091	$ & $	1.1E-07	$ & $	3.1E-07	$ & $	7.5E-07	$ & $	2.1E-06	$ & $	2.8E-06	$ & $	3.3E-06	$ & $	4.0E-06	$ & $	7.3E-07	$ & $	2.6E-06	$ & $	4.6E-07	$ \\
$	1087	$ & $	7.4E-08	$ & $	1.1E-07	$ & $	1.1E-06	$ & $	4.4E-06	$ & $	8.8E-06	$ & $	1.2E-05	$ & $	1.5E-05	$ & $	1.9E-07	$ & $	7.7E-06	$ & $	6.9E-08	$ \\
$	1045	$ & $	4.2E-07	$ & $	5.3E-07	$ & $	1.1E-06	$ & $	3.3E-06	$ & $	5.5E-06	$ & $	7.2E-06	$ & $	9.3E-06	$ & $	4.1E-06	$ & $	4.6E-06	$ & $	3.5E-06	$ \\
$	1109	$ & $	2.3E-07	$ & $	2.7E-07	$ & $	5.7E-07	$ & $	1.6E-06	$ & $	2.2E-06	$ & $	2.6E-06	$ & $	3.0E-06	$ & $	5.1E-07	$ & $	1.8E-06	$ & $	4.2E-08	$ \\
$	1002	$ & $	2.7E-07	$ & $	3.6E-07	$ & $	4.7E-07	$ & $	7.5E-07	$ & $	8.0E-07	$ & $	8.2E-07	$ & $	8.4E-07	$ & $	7.4E-07	$ & $	1.1E-07	$ & $	6.7E-08	$ \\
$	1061	$ & $	2.1E-07	$ & $	4.1E-07	$ & $	1.2E-06	$ & $	4.1E-06	$ & $	7.3E-06	$ & $	9.9E-06	$ & $	1.3E-05	$ & $	9.1E-07	$ & $	3.4E-06	$ & $	7.3E-07	$ \\
$	1022	$ & $	5.4E-09	$ & $	1.4E-08	$ & $	3.1E-08	$ & $	6.6E-08	$ & $	8.1E-08	$ & $	9.9E-08	$ & $	1.3E-07	$ & $	2.7E-08	$ & $	4.1E-08	$ & $	1.7E-08	$ \\
$	1000	$ & $	5.1E-08	$ & $	1.9E-07	$ & $	4.9E-07	$ & $	1.1E-06	$ & $	1.6E-06	$ & $	2.0E-06	$ & $	2.4E-06	$ & $	5.0E-07	$ & $	1.2E-06	$ & $	3.7E-07	$ \\
$	1020	$ & $	5.7E-07	$ & $	7.1E-07	$ & $	1.2E-06	$ & $	3.7E-06	$ & $	7.0E-06	$ & $	9.8E-06	$ & $	1.3E-05	$ & $	4.2E-06	$ & $	6.5E-06	$ & $	6.4E-07	$ \\
$	1081	$ & $	3.8E-07	$ & $	6.1E-07	$ & $	1.5E-06	$ & $	5.0E-06	$ & $	9.6E-06	$ & $	1.4E-05	$ & $	1.8E-05	$ & $	7.4E-06	$ & $	9.3E-06	$ & $	1.9E-06	$ \\
$	1074	$ & $	1.9E-07	$ & $	3.6E-07	$ & $	9.4E-07	$ & $	2.8E-06	$ & $	4.4E-06	$ & $	5.7E-06	$ & $	7.3E-06	$ & $	8.0E-08	$ & $	4.2E-06	$ & $	9.9E-10	$ \\
$	1027	$ & $	1.4E-07	$ & $	2.9E-07	$ & $	1.8E-06	$ & $	6.8E-06	$ & $	1.4E-05	$ & $	1.9E-05	$ & $	2.4E-05	$ & $	1.2E-05	$ & $	2.5E-06	$ & $	3.0E-06	$ \\
$	1062	$ & $	2.3E-08	$ & $	2.8E-07	$ & $	3.0E-06	$ & $	1.2E-05	$ & $	2.5E-05	$ & $	3.6E-05	$ & $	4.6E-05	$ & $	1.8E-05	$ & $	1.6E-05	$ & $	1.5E-05	$ \\
$	1077	$ & $	3.3E-07	$ & $	5.6E-07	$ & $	1.2E-06	$ & $	3.9E-06	$ & $	7.4E-06	$ & $	1.1E-05	$ & $	1.5E-05	$ & $	4.4E-06	$ & $	4.8E-06	$ & $	3.3E-06	$ \\
$	1028	$ & $	5.2E-07	$ & $	7.4E-07	$ & $	1.2E-06	$ & $	3.2E-06	$ & $	4.6E-06	$ & $	5.7E-06	$ & $	7.1E-06	$ & $	2.2E-06	$ & $	3.1E-06	$ & $	1.8E-06	$ \\
$	1092	$ & $	1.6E-07	$ & $	2.2E-07	$ & $	5.1E-07	$ & $	1.5E-06	$ & $	2.4E-06	$ & $	3.0E-06	$ & $	3.7E-06	$ & $	1.8E-06	$ & $	2.8E-06	$ & $	1.5E-06	$ \\
$	1012	$ & $	3.6E-07	$ & $	6.5E-07	$ & $	1.9E-06	$ & $	7.5E-06	$ & $	1.7E-05	$ & $	2.5E-05	$ & $	3.5E-05	$ & $	1.0E-05	$ & $	1.9E-05	$ & $	2.3E-06	$ \\
$	1040	$ & $	2.9E-07	$ & $	4.9E-07	$ & $	9.9E-07	$ & $	2.5E-06	$ & $	3.1E-06	$ & $	3.6E-06	$ & $	4.2E-06	$ & $	2.8E-06	$ & $	6.8E-07	$ & $	4.6E-07	$ \\
\hline
\end{tabular}}
\hfill{}
\end{minipage}
\end{table}

\begin{table}
\renewcommand\thetable{}
\centering
\begin{minipage}{220mm}
{\small
\hfill{}
\hspace*{-2.5cm}
\caption{\textbf{A6} Output from MAGPHYS: LSBs }
\begin{tabular}{|c|c|c|c|c|c|c|c|c|}
\hline
id   & $ M_{star}(M_{\odot})$  &sSFR$(yr^{-1})$ &SFR(0.1 Gyr)& $L_{dust}$  &$M_{dust}(M_{\odot})$  & $\gamma(Gyr^{-1})$  & N bursts & tform\\  
\hline
$	1048	$ & $	5.17E+08	$ & $	1.29E-09	$ & $	6.69E-01	$ & $	5.46E+09	$ & $	5.02E+06	$ & $	2.11E-01	$ & $	0.00E+00	$ & $	1.07E+09	$ \\
$	1084	$ & $	4.33E+07	$ & $	3.69E-10	$ & $	1.60E-02	$ & $	1.52E+08	$ & $	1.61E+05	$ & $	1.54E-01	$ & $	3.00E+00	$ & $	4.69E+09	$ \\
$	1141	$ & $	1.54E+10	$ & $	4.01E-11	$ & $	6.17E-01	$ & $	4.26E+09	$ & $	3.92E+06	$ & $	2.36E-01	$ & $	2.00E+00	$ & $	1.00E+10	$ \\
$	1010	$ & $	1.07E+10	$ & $	5.45E-11	$ & $	5.85E-01	$ & $	5.74E+09	$ & $	5.28E+06	$ & $	1.80E-02	$ & $	1.00E+00	$ & $	1.19E+10	$ \\
$	1025	$ & $	2.36E+09	$ & $	3.29E-11	$ & $	7.78E-02	$ & $	6.09E+08	$ & $	1.29E+06	$ & $	1.70E-02	$ & $	2.00E+00	$ & $	1.29E+10	$ \\
$	1085	$ & $	5.09E+08	$ & $	1.79E-10	$ & $	9.09E-02	$ & $	2.46E+08	$ & $	2.62E+05	$ & $	4.40E-01	$ & $	2.00E+00	$ & $	2.99E+09	$ \\
$	1093	$ & $	3.79E+08	$ & $	4.40E-11	$ & $	1.67E-02	$ & $	1.51E+09	$ & $	2.37E+06	$ & $	5.50E-01	$ & $	1.00E+00	$ & $	3.78E+09	$ \\
$	1013	$ & $	2.20E+09	$ & $	2.69E-10	$ & $	5.94E-01	$ & $	4.95E+09	$ & $	1.05E+07	$ & $	5.02E-01	$ & $	0.00E+00	$ & $	2.90E+09	$ \\
$	1039	$ & $	1.05E+10	$ & $	1.01E-10	$ & $	1.06E+00	$ & $	1.02E+09	$ & $	1.09E+06	$ & $	6.00E-02	$ & $	1.00E+00	$ & $	1.13E+10	$ \\
$	1181	$ & $	1.55E+10	$ & $	1.83E-10	$ & $	2.84E+00	$ & $	1.33E+10	$ & $	1.41E+07	$ & $	2.40E-02	$ & $	2.00E+00	$ & $	6.16E+09	$ \\
$	1094	$ & $	1.99E+08	$ & $	1.73E-10	$ & $	3.43E-02	$ & $	5.56E+08	$ & $	1.18E+06	$ & $	2.77E-01	$ & $	1.00E+00	$ & $	3.99E+09	$ \\
$	1142	$ & $	7.72E+09	$ & $	2.09E-10	$ & $	1.61E+00	$ & $	1.92E+10	$ & $	4.05E+07	$ & $	4.20E-02	$ & $	0.00E+00	$ & $	7.34E+09	$ \\
$	1071	$ & $	1.33E+10	$ & $	1.37E-10	$ & $	1.82E+00	$ & $	3.05E+10	$ & $	6.43E+07	$ & $	2.67E-01	$ & $	0.00E+00	$ & $	5.69E+09	$ \\
$	1133	$ & $	7.06E+08	$ & $	1.62E-11	$ & $	1.14E-02	$ & $	5.56E+07	$ & $	1.17E+05	$ & $	1.17E-01	$ & $	2.00E+00	$ & $	1.26E+10	$ \\
$	1057	$ & $	5.48E+09	$ & $	1.67E-10	$ & $	9.19E-01	$ & $	1.43E+10	$ & $	3.02E+07	$ & $	4.36E-01	$ & $	2.00E+00	$ & $	2.89E+09	$ \\
$	1097	$ & $	2.33E+10	$ & $	3.09E-11	$ & $	7.17E-01	$ & $	9.68E+09	$ & $	2.04E+07	$ & $	7.20E-02	$ & $	4.00E+00	$ & $	1.13E+10	$ \\
$	1116	$ & $	2.30E+09	$ & $	1.54E-10	$ & $	3.55E-01	$ & $	1.80E+09	$ & $	1.65E+06	$ & $	1.20E-01	$ & $	0.00E+00	$ & $	7.36E+09	$ \\
$	1083	$ & $	6.52E+08	$ & $	1.32E-09	$ & $	8.56E-01	$ & $	9.78E+09	$ & $	8.99E+06	$ & $	3.82E-01	$ & $	3.00E+00	$ & $	7.83E+09	$ \\
$	1179	$ & $	2.08E+09	$ & $	1.95E-10	$ & $	4.06E-01	$ & $	4.92E+09	$ & $	1.04E+07	$ & $	9.00E-02	$ & $	2.00E+00	$ & $	4.84E+09	$ \\
$	1155	$ & $	5.21E+09	$ & $	2.55E-10	$ & $	1.33E+00	$ & $	1.72E+10	$ & $	3.63E+07	$ & $	3.12E-01	$ & $	1.00E+00	$ & $	3.59E+09	$ \\
$	1098	$ & $	2.49E+10	$ & $	3.73E-11	$ & $	9.30E-01	$ & $	7.05E+09	$ & $	6.48E+06	$ & $	2.02E-01	$ & $	1.00E+00	$ & $	1.12E+10	$ \\
$	1110	$ & $	1.23E+09	$ & $	4.71E-11	$ & $	5.78E-02	$ & $	2.79E+08	$ & $	2.57E+05	$ & $	8.90E-02	$ & $	2.00E+00	$ & $	1.20E+10	$ \\
$	1178	$ & $	1.57E+10	$ & $	7.74E-11	$ & $	1.21E+00	$ & $	5.34E+09	$ & $	4.91E+06	$ & $	1.54E-01	$ & $	4.00E+00	$ & $	8.14E+09	$ \\
$	1143	$ & $	1.14E+09	$ & $	1.87E-10	$ & $	2.13E-01	$ & $	9.01E+08	$ & $	9.58E+05	$ & $	3.00E-03	$ & $	0.00E+00	$ & $	9.59E+09	$ \\
$	1024	$ & $	1.11E+10	$ & $	1.77E-10	$ & $	1.95E+00	$ & $	1.57E+10	$ & $	1.44E+07	$ & $	1.18E-01	$ & $	0.00E+00	$ & $	6.71E+09	$ \\
$	1159	$ & $	1.23E+10	$ & $	1.96E-10	$ & $	2.41E+00	$ & $	3.20E+10	$ & $	6.76E+07	$ & $	1.20E-01	$ & $	1.00E+00	$ & $	4.01E+09	$ \\
$	1101	$ & $	2.45E+08	$ & $	1.52E-10	$ & $	3.72E-02	$ & $	2.53E+08	$ & $	2.32E+05	$ & $	4.60E-02	$ & $	1.00E+00	$ & $	8.66E+09	$ \\
$	1050	$ & $	2.34E+10	$ & $	1.07E-10	$ & $	2.51E+00	$ & $	2.84E+10	$ & $	5.99E+07	$ & $	3.80E-01	$ & $	2.00E+00	$ & $	3.43E+09	$ \\
$	1162	$ & $	1.08E+10	$ & $	4.30E-11	$ & $	4.63E-01	$ & $	9.28E+09	$ & $	1.96E+07	$ & $	5.76E-01	$ & $	0.00E+00	$ & $	5.72E+09	$ \\
$	1038	$ & $	9.19E+09	$ & $	1.54E-11	$ & $	1.41E-01	$ & $	6.85E+09	$ & $	6.30E+06	$ & $	1.56E-01	$ & $	5.00E+00	$ & $	1.32E+10	$ \\
$	1148	$ & $	1.36E+10	$ & $	6.82E-11	$ & $	9.29E-01	$ & $	7.93E+09	$ & $	7.29E+06	$ & $	2.60E-02	$ & $	2.00E+00	$ & $	8.46E+09	$ \\
$	1032	$ & $	8.37E+09	$ & $	4.16E-11	$ & $	3.48E-01	$ & $	1.38E+09	$ & $	1.27E+06	$ & $	2.00E-03	$ & $	3.00E+00	$ & $	8.91E+09	$ \\
$	1125	$ & $	9.87E+09	$ & $	9.78E-11	$ & $	9.68E-01	$ & $	8.11E+09	$ & $	7.46E+06	$ & $	2.21E-01	$ & $	1.00E+00	$ & $	2.80E+09	$ \\
$	1154	$ & $	8.02E+09	$ & $	4.49E-11	$ & $	3.60E-01	$ & $	4.80E+09	$ & $	1.02E+07	$ & $	5.27E-01	$ & $	2.00E+00	$ & $	4.55E+09	$ \\
$	1158	$ & $	3.52E+10	$ & $	6.12E-11	$ & $	2.16E+00	$ & $	1.97E+10	$ & $	1.81E+07	$ & $	7.70E-02	$ & $	1.00E+00	$ & $	6.12E+09	$ \\
$	1117	$ & $	3.52E+09	$ & $	3.76E-10	$ & $	1.32E+00	$ & $	7.75E+09	$ & $	7.12E+06	$ & $	1.00E-01	$ & $	1.00E+00	$ & $	3.36E+09	$ \\
$	1008	$ & $	5.21E+09	$ & $	4.83E-11	$ & $	2.52E-01	$ & $	3.16E+09	$ & $	6.68E+06	$ & $	1.77E-01	$ & $	3.00E+00	$ & $	5.19E+09	$ \\
$	1014	$ & $	1.31E+10	$ & $	1.01E-10	$ & $	1.31E+00	$ & $	1.27E+09	$ & $	1.35E+06	$ & $	6.00E-02	$ & $	1.00E+00	$ & $	1.13E+10	$ \\
$	1090	$ & $	3.08E+09	$ & $	2.23E-10	$ & $	6.85E-01	$ & $	4.14E+09	$ & $	3.81E+06	$ & $	2.60E-01	$ & $	0.00E+00	$ & $	4.31E+09	$ \\
$	1066	$ & $	6.85E+09	$ & $	3.55E-11	$ & $	2.44E-01	$ & $	5.03E+09	$ & $	1.06E+07	$ & $	3.70E-02	$ & $	3.00E+00	$ & $	1.25E+10	$ \\
$	1106	$ & $	1.72E+09	$ & $	3.10E-10	$ & $	5.35E-01	$ & $	7.91E+09	$ & $	1.67E+07	$ & $	7.90E-02	$ & $	0.00E+00	$ & $	4.76E+09	$ \\
$	1076	$ & $	9.33E+09	$ & $	3.95E-11	$ & $	3.69E-01	$ & $	4.14E+09	$ & $	8.75E+06	$ & $	1.07E-01	$ & $	3.00E+00	$ & $	8.20E+09	$ \\
$	1170	$ & $	1.10E+10	$ & $	1.90E-10	$ & $	2.09E+00	$ & $	8.10E+09	$ & $	7.45E+06	$ & $	8.50E-02	$ & $	3.00E+00	$ & $	5.80E+09	$ \\
$	1001	$ & $	2.08E+10	$ & $	2.86E-11	$ & $	5.96E-01	$ & $	1.07E+10	$ & $	2.25E+07	$ & $	4.10E-01	$ & $	0.00E+00	$ & $	8.20E+09	$ \\
$	1118	$ & $	2.55E+09	$ & $	2.30E-11	$ & $	5.87E-02	$ & $	1.98E+09	$ & $	2.11E+06	$ & $	5.04E-01	$ & $	3.00E+00	$ & $	1.30E+10	$ \\
$	1064	$ & $	3.33E+10	$ & $	6.67E-11	$ & $	2.22E+00	$ & $	2.08E+10	$ & $	3.94E+07	$ & $	4.70E-02	$ & $	4.00E+00	$ & $	1.19E+10	$ \\
$	1124	$ & $	7.92E+07	$ & $	2.74E-10	$ & $	2.17E-02	$ & $	2.70E+08	$ & $	5.71E+05	$ & $	6.10E-02	$ & $	1.00E+00	$ & $	4.50E+09	$ \\
$	1134	$ & $	2.15E+09	$ & $	4.53E-11	$ & $	9.73E-02	$ & $	3.63E+08	$ & $	3.34E+05	$ & $	1.43E-01	$ & $	1.00E+00	$ & $	8.08E+09	$ \\
$	1089	$ & $	2.95E+09	$ & $	3.52E-10	$ & $	1.04E+00	$ & $	6.51E+09	$ & $	5.99E+06	$ & $	6.33E-01	$ & $	1.00E+00	$ & $	3.95E+09	$ \\
$	1163	$ & $	3.26E+09	$ & $	2.25E-10	$ & $	7.36E-01	$ & $	8.44E+09	$ & $	1.78E+07	$ & $	1.70E-01	$ & $	2.00E+00	$ & $	2.44E+09	$ \\
$	1166	$ & $	2.56E+09	$ & $	1.40E-10	$ & $	3.58E-01	$ & $	3.96E+09	$ & $	8.37E+06	$ & $	4.68E-01	$ & $	1.00E+00	$ & $	4.10E+09	$ \\
$	1088	$ & $	3.74E+09	$ & $	1.09E-10	$ & $	4.07E-01	$ & $	8.20E+08	$ & $	7.54E+05	$ & $	1.64E-01	$ & $	0.00E+00	$ & $	8.12E+09	$ \\
$	1152	$ & $	7.14E+10	$ & $	1.14E-12	$ & $	8.11E-02	$ & $	6.12E+09	$ & $	9.57E+06	$ & $	6.68E-01	$ & $	1.00E+00	$ & $	9.35E+09	$ \\
$	1063	$ & $	2.21E+10	$ & $	8.39E-11	$ & $	1.86E+00	$ & $	1.45E+10	$ & $	1.34E+07	$ & $	3.23E-01	$ & $	1.00E+00	$ & $	5.88E+09	$ \\
$	1069	$ & $	8.40E+09	$ & $	2.16E-10	$ & $	1.80E+00	$ & $	1.81E+10	$ & $	3.81E+07	$ & $	5.37E-01	$ & $	0.00E+00	$ & $	3.15E+09	$ \\
$	1102	$ & $	3.12E+10	$ & $	1.86E-11	$ & $	5.82E-01	$ & $	3.14E+09	$ & $	2.89E+06	$ & $	2.17E-01	$ & $	2.00E+00	$ & $	1.19E+10	$ \\
$	1147	$ & $	2.06E+10	$ & $	1.33E-10	$ & $	2.75E+00	$ & $	2.74E+10	$ & $	5.80E+07	$ & $	1.43E-01	$ & $	0.00E+00	$ & $	7.62E+09	$ \\
$	1135	$ & $	3.18E+09	$ & $	7.74E-11	$ & $	2.46E-01	$ & $	2.32E+09	$ & $	4.90E+06	$ & $	2.43E-01	$ & $	4.00E+00	$ & $	6.39E+09	$ \\
$	1130	$ & $	3.65E+08	$ & $	5.58E-10	$ & $	2.03E-01	$ & $	2.77E+09	$ & $	5.85E+06	$ & $	1.73E-01	$ & $	0.00E+00	$ & $	2.47E+09	$ \\
$	1044	$ & $	5.97E+09	$ & $	1.47E-10	$ & $	8.74E-01	$ & $	1.05E+10	$ & $	2.22E+07	$ & $	5.10E-02	$ & $	2.00E+00	$ & $	7.68E+09	$ \\
$	1119	$ & $	5.23E+10	$ & $	3.71E-11	$ & $	1.94E+00	$ & $	6.81E+09	$ & $	6.26E+06	$ & $	1.80E-01	$ & $	2.00E+00	$ & $	1.04E+10	$ \\
$	1016	$ & $	3.42E+09	$ & $	2.00E-10	$ & $	6.86E-01	$ & $	5.39E+09	$ & $	4.95E+06	$ & $	6.90E-02	$ & $	1.00E+00	$ & $	6.32E+09	$ \\
$	1120	$ & $	4.34E+08	$ & $	6.13E-10	$ & $	2.66E-01	$ & $	1.81E+09	$ & $	1.67E+06	$ & $	5.04E-01	$ & $	2.00E+00	$ & $	8.41E+09	$ \\
$	1046	$ & $	6.58E+09	$ & $	9.47E-13	$ & $	6.24E-03	$ & $	8.10E+08	$ & $	1.27E+06	$ & $	6.10E-01	$ & $	2.00E+00	$ & $	1.12E+10	$ \\
\hline
\end{tabular}}
\hfill{}
\end{minipage}
\end{table}

\begin{table}
\renewcommand\thetable{}
\centering
\begin{minipage}{220mm}
{\small
\hfill{}
\hspace*{-2.5cm}
\caption{\textbf{A7} Output from MAGPHYS: STs}
\begin{tabular}{|c|c|c|c|c|c|c|c|c|}
\hline
id  \footnote{Galaxy ID from \protect \cite{refId0}}& $ M_{star}(M_{\odot})$ \footnote{Stellar mass ($M_{\odot}$)} &sSFR \footnote{Specific star formation rate ($yr^{-1}$)} &SFR (0.1 Gyr) \footnote{Star formation rate averaged over last 0.1 Gyr}& $L_{dust}$  \footnote{luminosity of dust {$L_{\odot}$}}&$M_{dust}(M_{\odot})$ \footnote{Dust mass ($M_{\odot}$)} & $\gamma(Gyr^{-1})$ \footnote{Star formation time scale} & N bursts \footnote{No of bursts since tform}& tform \footnote{Age of the oldest population (Gyr)} \\
\hline
$	1115	$ & $	1.58E+10	$ & $	3.42E-11	$ & $	5.39E-01	$ & $	7.80E+09	$ & $	1.65E+07	$ & $	1.19E-01	$ & $	4.00E+00	$ & $	1.26E+10	 $ \\
$	1076	$ & $	8.99E+09	$ & $	2.96E-11	$ & $	2.66E-01	$ & $	3.88E+09	$ & $	8.19E+06	$ & $	3.01E-01	$ & $	1.00E+00	$ & $	7.26E+09	 $ \\
$	1001	$ & $	7.37E+09	$ & $	7.85E-11	$ & $	5.79E-01	$ & $	6.91E+09	$ & $	1.46E+07	$ & $	1.16E-01	$ & $	3.00E+00	$ & $	1.03E+10	 $ \\
$	1118	$ & $	1.10E+10	$ & $	2.18E-11	$ & $	2.40E-01	$ & $	6.51E+09	$ & $	1.10E+07	$ & $	3.96E-01	$ & $	1.00E+00	$ & $	9.05E+09	 $ \\
$	1064	$ & $	2.51E+10	$ & $	1.51E-11	$ & $	3.79E-01	$ & $	1.62E+09	$ & $	1.49E+06	$ & $	3.10E-01	$ & $	4.00E+00	$ & $	1.08E+10	 $ \\
$	1047	$ & $	5.00E+09	$ & $	8.61E-11	$ & $	4.30E-01	$ & $	8.01E+09	$ & $	1.69E+07	$ & $	2.01E-01	$ & $	2.00E+00	$ & $	3.94E+09	 $ \\
$	1030	$ & $	2.21E+10	$ & $	1.37E-11	$ & $	3.02E-01	$ & $	6.02E+09	$ & $	1.27E+07	$ & $	5.07E-01	$ & $	2.00E+00	$ & $	6.14E+09	 $ \\
$	1089	$ & $	8.70E+09	$ & $	4.66E-11	$ & $	4.06E-01	$ & $	7.39E+09	$ & $	1.56E+07	$ & $	3.03E-01	$ & $	1.00E+00	$ & $	8.41E+09	 $ \\
$	1006	$ & $	6.90E+09	$ & $	1.59E-12	$ & $	1.09E-02	$ & $	7.85E+09	$ & $	5.16E+07	$ & $	5.87E-01	$ & $	2.00E+00	$ & $	9.70E+09	 $ \\
$	1063	$ & $	1.16E+10	$ & $	4.41E-11	$ & $	5.11E-01	$ & $	1.03E+10	$ & $	2.18E+07	$ & $	2.46E-01	$ & $	1.00E+00	$ & $	9.84E+09	 $ \\
$	1069	$ & $	2.04E+10	$ & $	4.69E-11	$ & $	9.54E-01	$ & $	1.16E+10	$ & $	2.45E+07	$ & $	2.57E-01	$ & $	1.00E+00	$ & $	9.37E+09	 $ \\
$	1049	$ & $	3.56E+09	$ & $	6.57E-10	$ & $	2.34E+00	$ & $	2.44E+10	$ & $	4.63E+07	$ & $	1.98E-01	$ & $	0.00E+00	$ & $	2.03E+09	 $ \\
$	1102	$ & $	5.00E+09	$ & $	5.24E-11	$ & $	2.62E-01	$ & $	1.26E+09	$ & $	1.16E+06	$ & $	2.90E-01	$ & $	1.00E+00	$ & $	4.14E+09	 $ \\
$	1103	$ & $	1.31E+10	$ & $	2.31E-11	$ & $	3.02E-01	$ & $	5.06E+09	$ & $	1.07E+07	$ & $	3.33E-01	$ & $	1.00E+00	$ & $	1.01E+10	 $ \\
$	1044	$ & $	1.37E+10	$ & $	3.82E-11	$ & $	5.23E-01	$ & $	8.49E+09	$ & $	1.79E+07	$ & $	1.50E-02	$ & $	4.00E+00	$ & $	9.62E+09	 $ \\
$	1119	$ & $	8.28E+08	$ & $	7.07E-12	$ & $	5.84E-03	$ & $	6.97E+07	$ & $	1.47E+05	$ & $	2.89E-01	$ & $	2.00E+00	$ & $	1.24E+10	 $ \\
$	1029	$ & $	1.03E+10	$ & $	4.18E-11	$ & $	4.29E-01	$ & $	5.11E+09	$ & $	1.08E+07	$ & $	1.46E-01	$ & $	2.00E+00	$ & $	1.31E+10	 $ \\
$	1046	$ & $	1.83E+10	$ & $	3.67E-11	$ & $	6.73E-01	$ & $	1.61E+10	$ & $	2.72E+07	$ & $	2.16E-01	$ & $	2.00E+00	$ & $	9.86E+09	 $ \\
$	1018	$ & $	9.65E+08	$ & $	6.21E-10	$ & $	5.99E-01	$ & $	8.45E+09	$ & $	1.79E+07	$ & $	7.95E-01	$ & $	0.00E+00	$ & $	1.46E+09	 $ \\
$	1042	$ & $	7.04E+09	$ & $	3.29E-11	$ & $	2.32E-01	$ & $	1.82E+09	$ & $	3.84E+06	$ & $	1.70E-02	$ & $	2.00E+00	$ & $	1.29E+10	 $ \\
$	1113	$ & $	3.93E+09	$ & $	1.86E-10	$ & $	7.32E-01	$ & $	9.51E+09	$ & $	2.01E+07	$ & $	2.00E-02	$ & $	2.00E+00	$ & $	7.67E+09	 $ \\
$	1053	$ & $	2.55E+09	$ & $	1.23E-10	$ & $	3.14E-01	$ & $	2.11E+09	$ & $	1.94E+06	$ & $	8.10E-02	$ & $	2.00E+00	$ & $	2.02E+09	 $ \\
$	1015	$ & $	3.39E+09	$ & $	5.13E-11	$ & $	1.74E-01	$ & $	1.65E+09	$ & $	3.49E+06	$ & $	3.00E-03	$ & $	3.00E+00	$ & $	8.99E+09	 $ \\
$	1075	$ & $	7.82E+09	$ & $	2.06E-12	$ & $	1.61E-02	$ & $	4.22E+08	$ & $	7.09E+05	$ & $	5.85E-01	$ & $	2.00E+00	$ & $	7.96E+09	 $ \\
$	1031	$ & $	1.71E+10	$ & $	5.74E-11	$ & $	9.81E-01	$ & $	1.84E+10	$ & $	3.88E+07	$ & $	1.94E-01	$ & $	1.00E+00	$ & $	5.16E+09	 $ \\
$	1036	$ & $	4.76E+08	$ & $	1.23E-09	$ & $	5.84E-01	$ & $	6.16E+09	$ & $	5.66E+06	$ & $	6.80E-02	$ & $	0.00E+00	$ & $	1.21E+09	 $ \\
$	1104	$ & $	4.86E+08	$ & $	3.02E-10	$ & $	1.47E-01	$ & $	1.12E+09	$ & $	1.03E+06	$ & $	1.55E-01	$ & $	0.00E+00	$ & $	4.13E+09	 $ \\
$	1091	$ & $	1.04E+09	$ & $	6.43E-09	$ & $	6.72E+00	$ & $	1.96E+10	$ & $	1.29E+08	$ & $	2.53E-01	$ & $	3.00E+00	$ & $	6.16E+09	 $ \\
$	1087	$ & $	7.43E+09	$ & $	3.13E-12	$ & $	2.33E-02	$ & $	5.63E+07	$ & $	1.19E+05	$ & $	3.87E-01	$ & $	3.00E+00	$ & $	1.00E+10	 $ \\
$	1045	$ & $	5.55E+09	$ & $	7.47E-11	$ & $	4.15E-01	$ & $	3.14E+09	$ & $	2.89E+06	$ & $	1.23E-01	$ & $	1.00E+00	$ & $	1.11E+10	 $ \\
$	1109	$ & $	6.07E+08	$ & $	1.22E-10	$ & $	7.39E-02	$ & $	1.13E+08	$ & $	1.04E+05	$ & $	1.53E-01	$ & $	1.00E+00	$ & $	2.70E+09	 $ \\
$	1002	$ & $	6.82E+07	$ & $	1.20E-08	$ & $	8.19E-01	$ & $	8.10E+09	$ & $	7.44E+06	$ & $	5.47E-01	$ & $	0.00E+00	$ & $	1.06E+08	 $ \\
$	1061	$ & $	7.13E+09	$ & $	1.07E-10	$ & $	7.61E-01	$ & $	1.10E+10	$ & $	2.32E+07	$ & $	5.20E-01	$ & $	0.00E+00	$ & $	4.41E+09	 $ \\
$	1022	$ & $	2.08E+07	$ & $	2.99E-09	$ & $	6.21E-02	$ & $	6.03E+08	$ & $	5.55E+05	$ & $	2.85E-01	$ & $	0.00E+00	$ & $	4.59E+08	 $ \\
$	1000	$ & $	7.95E+08	$ & $	1.32E-10	$ & $	1.05E-01	$ & $	6.59E+09	$ & $	7.02E+06	$ & $	9.52E-01	$ & $	5.00E+00	$ & $	1.21E+10	 $ \\
$	1020	$ & $	1.21E+10	$ & $	2.72E-11	$ & $	3.29E-01	$ & $	2.13E+09	$ & $	4.50E+06	$ & $	9.00E-02	$ & $	3.00E+00	$ & $	1.22E+10	 $ \\
$	1081	$ & $	1.68E+10	$ & $	3.17E-11	$ & $	5.32E-01	$ & $	5.28E+09	$ & $	4.85E+06	$ & $	1.67E-01	$ & $	3.00E+00	$ & $	1.23E+10	 $ \\
$	1074	$ & $	2.44E+09	$ & $	3.16E-10	$ & $	7.70E-01	$ & $	8.63E+09	$ & $	1.82E+07	$ & $	7.09E-01	$ & $	0.00E+00	$ & $	2.24E+09	 $ \\
$	1027	$ & $	2.13E+10	$ & $	1.45E-11	$ & $	3.08E-01	$ & $	1.12E+10	$ & $	1.75E+07	$ & $	4.44E-01	$ & $	0.00E+00	$ & $	9.30E+09	 $ \\
$	1062	$ & $	6.53E+10	$ & $	1.91E-12	$ & $	1.24E-01	$ & $	4.09E+10	$ & $	2.69E+08	$ & $	7.74E-01	$ & $	0.00E+00	$ & $	8.70E+09	 $ \\
$	1077	$ & $	1.40E+10	$ & $	3.75E-11	$ & $	5.24E-01	$ & $	5.77E+09	$ & $	1.22E+07	$ & $	1.51E-01	$ & $	1.00E+00	$ & $	1.26E+10	 $ \\
$	1028	$ & $	2.55E+09	$ & $	1.43E-10	$ & $	3.63E-01	$ & $	2.32E+09	$ & $	2.13E+06	$ & $	5.10E-02	$ & $	2.00E+00	$ & $	6.27E+09	 $ \\
$	1092	$ & $	1.43E+09	$ & $	1.44E-10	$ & $	2.06E-01	$ & $	2.15E+09	$ & $	1.98E+06	$ & $	1.93E-01	$ & $	1.00E+00	$ & $	4.54E+09	 $ \\
$	1012	$ & $	3.98E+10	$ & $	1.82E-11	$ & $	7.22E-01	$ & $	1.31E+10	$ & $	2.76E+07	$ & $	2.70E-01	$ & $	1.00E+00	$ & $	1.10E+10	 $ \\
$	1040	$ & $	8.72E+08	$ & $	9.80E-10	$ & $	8.55E-01	$ & $	1.14E+10	$ & $	2.42E+07	$ & $	5.02E-01	$ & $	0.00E+00	$ & $	1.21E+09	 $ \\

\hline
\end{tabular}}
\hfill{}
\end{minipage}
\end{table}

\end{document}